\documentclass{IEEEtran}
\IEEEoverridecommandlockouts
\usepackage{acronym}
\acrodef{6g}[6G]{six-generation}
\acrodef{mimo}[MIMO]{multi-input multi-output}
\acrodef{siso}[SISO]{single-input single-output} 
\usepackage{amsmath}
\usepackage{graphicx} 
\usepackage{epstopdf}
\usepackage{amssymb}
\usepackage{amsfonts}
\usepackage{booktabs} 
\usepackage{amsthm}
\usepackage{cite}
\usepackage{relsize}
\usepackage{bm,comment}
\usepackage{algorithm}
\usepackage{algorithmic}

\usepackage{subeqnarray}
\usepackage{subfigure}
\usepackage{multicol}
\usepackage{multirow}
\usepackage{diagbox}
\usepackage{slashbox}
\usepackage{stfloats}
\usepackage{float}
\usepackage{color}
\usepackage{cases}
\usepackage{CJK}
\usepackage{bm}

\usepackage[colorlinks,linkcolor=blue,citecolor=red]{hyperref}
\usepackage{lipsum}

\usepackage{geometry}
 \def\BibTeX{{\rm B\kern-.05em{\sc i\kern-.025em b}\kern-.08em
     T\kern-.1667em\lower.7ex\hbox{E}\kern-.125emX}}

\begin{document}
\title{Hybrid-Field Interference Suppression for Low-Complexity Uplink Channel Estimation}

\author{
Zihan Zhou,
Li Chen,~\IEEEmembership{Senior Member,~IEEE}, Ang Chen, Weidong Wang,\\ Changsheng You,~\IEEEmembership{ Member,~IEEE}, and 
Yunfei Chen,~\IEEEmembership{Fellow,~IEEE}
\vspace{-0.5cm}

\thanks{
Zihan Zhou, Li Chen, Ang Chen, and Weidong Wang are with the CAS Key Laboratory of Wireless Optical Communication, University of Science and Technology of China (USTC), Hefei 230027, China (e-mail: zzh31@mail.ustc.edu.cn; chenli87@ustc.edu.cn; chenang1122@mail.ustc.edu.cn; wdwang@ustc.edu.cn).

Changsheng You is with the Department of Electronic and Electrical Engineering, Southern University of Science and Technology (SUSTech), Shenzhen 518055, China (e-mail: youcs@sustech.edu.cn).

Yunfei Chen is with the Department of Engineering, University of Durham, DH1 3LE Durham, U.K. (e-mail: yunfei.chen@durham.ac.uk).
}
}

\maketitle
\begin{abstract}
Uplink channel estimation can be severely degraded by hybrid-field interference, where strong near-field intra-site interference is coupled with far-field inter-site interference. 
Although a joint near-/far-field dictionary can capture this interference structure, it requires dense angle--distance matching that may incur high complexity and performance degradation.
To address these challenges, this paper proposes a hybrid-field interference suppression framework for uplink channel estimation. First, we formulate a hybrid-field interference model and analyze its sparse reconstruction. Then, to avoid exhaustive near-field dictionary matching, we propose a low-complexity interference-parameter estimation method based on the angle--domain diffusion pattern of the dominant interference.
Using the estimated parameters, the proposed parametric interference-nulling orthogonal matching pursuit (PIN-OMP) algorithm suppresses the dominant interference components and reconstructs the desired uplink channel from the interference-suppressed observation.
Furthermore, a physical critical-distance bound is derived to characterize the feasible distance-inversion region, and a conditional hybrid Cram\'er--Rao bound (HCRB) is established to quantify the desired channel reconstruction limit under a reconstructed interference basis. Finally, numerical results are provided to demonstrate the effectiveness of the proposed scheme.
\end{abstract}


\section{Introduction}

\IEEEPARstart{I}{nterference} has long been a fundamental problem in wireless communication systems. In beyond-5G and future 6G networks, large-aperture antenna arrays, high-frequency bands, and dense deployments are expected to improve spectral efficiency and uplink reception capability, while also increasing the interference at the receiver~\cite{lu2024tutorial,dai2026tutorial,zhang2022beamfocusing}.  This trend is particularly critical for uplink communication, where a weak desired signal may coexist with strong unintended transmissions from nearby cells or remote transmitters. Thus, suppressing dominant interference remains essential for reliable uplink signal recovery in emerging large-array wireless systems~\cite{shen2012dynamic}.

Several studies have addressed interference suppression for channel estimation in the far field. Specifically, the work in~\cite{sun2020ictsp} showed that conventional least-squares (LS) estimators exploit temporal pilot correlation, but may become interference-limited under high interference-to-noise ratio (INR) with finite orthogonal pilots. An interference-aware minimum mean-square error (MMSE) receiver was proposed in~\cite{wrulich2010mmse} to exploit the spatial structure and covariance of interference. Nonetheless, its performance depends on accurate interference covariance estimation, which may require substantial pilot observations in large-array systems. To reduce pilot overhead, sparse methods have been introduced by exploiting angular sparsity, where the received signal is represented over a far-field angular dictionary and dominant components are recovered by sparse recovery techniques~\cite{cotter2002sparse,lee2016omp,tang2013offgrid}. These methods are effective when the received wavefronts can be well approximated by plane waves. Nevertheless, with the emergence of extremely large-scale arrays (XL-arrays), nearby interferers may enter the radiating near-field region, where spherical-wave propagation weakens the one-dimensional (1D) angular sparsity and degrades far-field dictionary representations~\cite{han2020xlmimo}.

To characterize and suppress near-field interference, recent studies have investigated near-field channel modeling, sparse representation, and interference management. Unlike far-field interference, which is mainly separated by angular resolution, near-field interference is jointly determined by angle and distance, and its energy may spread over a wide angular region when projected onto a far-field dictionary.  Therefore, its wavefront should be modeled as spherical, and the phase of the near-field array response becomes a nonlinear function of the antenna index~\cite{cui2022farnear}. To capture this structure, polar-domain or angle--distance dictionaries were introduced, where each near-field atom was indexed by both direction and range~\cite{zhang2024nfsparse,wu2024dft}.
 In addition, near-field multiuser studies have shown that beam focusing can exploit the distance dimension to distinguish users sharing similar angular directions, thereby reducing inter-user interference beyond conventional far-field angular separation~\cite{wu2023ldma}. These results indicate that near-field propagation provides additional spatial degrees of freedom, but also requires more accurate angle--distance modeling for interference reconstruction and suppression.

In practical large-aperture uplink systems, the interference often exhibits a hybrid-field structure. Specifically, nearby co-sited sectors and strong local transmitters can generate near-field intra-site interference with spherical-wave characteristics, whereas remote base stations (BSs) or transmitters can produce far-field inter-site interference with plane-wave angular responses~\cite{long2018interference}. Recently, hybrid-field channel modeling and interference analysis
have attracted increasing attention. For example, the works
in~\cite{xiao_mixed_targets,xiao2026movable} investigated the impact of array geometry on hybrid-field system performance. Hybrid-field channel estimation is further introduced with joint near-/far-field dictionaries to represent spherical-wave and plane-wave components in a unified form~\cite{yue2024hybrid}. Moreover, efficient near-field feature estimation and discrete Fourier transform (DFT)-codebook-based beam training demonstrated that near-field angle and distance can be inferred with reduced search complexity when the source distance is sufficiently large or when structured amplitude-domain features are available~\cite{chung2025efficient,wang2026precise}. These studies provide useful insights into hybrid-field representation and low-complexity parameter inference.

However, under high INR, strong structured hybrid-field interference can dominate the pilot-correlated observation and mask the weak desired component, making direct channel recovery highly sensitive to residual interference. While hybrid-field channel modeling and parameter analysis have attracted increasing attention, few studies have addressed hybrid-field interference suppression for uplink channel estimation under such interference-dominated conditions. A key challenge is to suppress hybrid-field interference
components with distinct spatial signatures. Specifically, dominant near-field intra-site interference is typically governed by quasi-static geometric parameters over a pilot block and exhibits structured angular diffusion after angle--domain projection, while far-field interference remains angularly sparse.

To address this challenge, in this paper, we develop an uplink channel estimation framework under strong hybrid-field interference. First, the proposed framework reconstructs the dominant hybrid-field interference based on the received signal. Then, the resulting interference subspace is nulled before the desired uplink channel is reconstructed from the interference-suppressed observation. To avoid exhaustive angle--distance dictionary matching, we propose a low-complexity interference-parameter estimation method. Specifically, the structured angular diffusion of the dominant near-field interference in the 1D far-field angular spectrum is exploited to estimate the interference angle from the diffusion-envelope symmetry and estimate the interference distance through monotone power--curvature inversion. Based on the estimated interference parameters, we further develop a parametric interference-nulling orthogonal matching pursuit (PIN-OMP) algorithm for hybrid-field interference suppression and desired uplink channel reconstruction.  The main contributions of this paper are summarized below: 

\begin{itemize}
    \item \textbf{Hybrid-field interference modeling and sparse reconstruction framework:}
    We formulate an uplink communication model under hybrid-field interference, where the desired uplink channel and far-field inter-site interference are modeled by plane-wave responses, while near-field intra-site interference is represented by spherical-wave responses. Based on this model, we develop a dictionary-based sparse reconstruction framework to reconstruct dominant hybrid-field interference, null the reconstructed interference subspace, and reconstruct the desired uplink channel from the interference-suppressed observation. This framework also identifies dense near-field angle--distance matching as a key bottleneck in dictionary-based hybrid-field interference reconstruction.

    \item \textbf{Low-complexity interference parameter estimation and PIN-OMP design:}
To avoid exhaustive near-field angle--distance dictionary matching, we exploit the angular diffusion of the dominant near-field intra-site interference in the 1D far-field angular spectrum. Specifically, the interference angle is estimated from the symmetry of the diffusion envelope, while the interference distance is estimated through monotone power--curvature inversion implemented by piecewise cubic Hermite interpolating polynomial (PCHIP) interpolation. Using the estimated parameters, we propose the PIN-OMP algorithm, which reconstructs and nulls the near-field intra-site interference basis, reconstructs residual far-field inter-site interference, and recovers the desired uplink channel over the projected far-field dictionary.

    \item \textbf{Feasibility and performance bounds with complexity analysis:}
To characterize the reliability of the proposed framework, we analyze PIN-OMP from the perspectives of distance-inversion feasibility, desired-channel reconstruction limit, and computational complexity. First, the monotone distance-inversion range is translated into a physical critical-distance bound, which specifies the feasible operating region of the proposed parameter-estimation method. Second, a conditional hybrid Cram\'er--Rao bound (HCRB) is derived to quantify the desired channel reconstruction limit under a reconstructed interference basis. Third, the dominant computational complexity is analyzed to show that PIN-OMP avoids exhaustive near-field angle--distance matching. Numerical results validate the effectiveness of PIN-OMP under different SNRs, pilot lengths, and intra-site INR values.
\end{itemize}

The rest of this paper is organized as follows: Section~\ref{sec2} presents the system model and the limitations of conventional estimators. Section~\ref{sec3} analyzes a dictionary-based framework for hybrid-field interference suppression. Section~\ref{sec4} develops a low-complexity interference-parameter estimation method for hybrid-field interference reconstruction. Section~\ref{sec5} presents PIN-OMP and provides performance and complexity analyses. Section~\ref{sec6} gives numerical results, and Section~\ref{sec7} concludes the paper.

\emph{Notations:} 
In this paper, vectors and matrices are denoted by lowercase and uppercase boldface letters, respectively. We use superscripts $(\cdot)^{T}$, $(\cdot)^{-1}$, and $(\cdot)^{H}$ to denote the transpose, inverse, and Hermitian transpose, respectively. The symbol $\|\cdot\|_2$ denotes the $\ell_2$-norm, and $|\cdot|$ represents the modulus operator. $\mathbb{C}^{M \times N}$ is the set of all $M \times N$ complex matrices, and $\mathbb{E}[\cdot]$ denotes the mathematical expectation. The superscript \((\cdot)^\dagger\) denotes the Moore--Penrose pseudoinverse, and \(\operatorname{tr}(\cdot)\) denotes the trace. $\mathbf{I}_N$ denotes the $N \times N$ identity matrix, $\mathbf{0}$ is a zero matrix. The imaginary unit is denoted by $j = \sqrt{-1}$, and $\mathcal{CN}\left(\mu, \sigma^2\right)$ denotes the circularly symmetric complex Gaussian distribution with mean $\mu$ and variance $\sigma^2$.

\section{System Model and Problem Formulation}\label{sec2}

\subsection{System Model}
\begin{figure}[t]
    \centering
\includegraphics[width=0.95\linewidth]{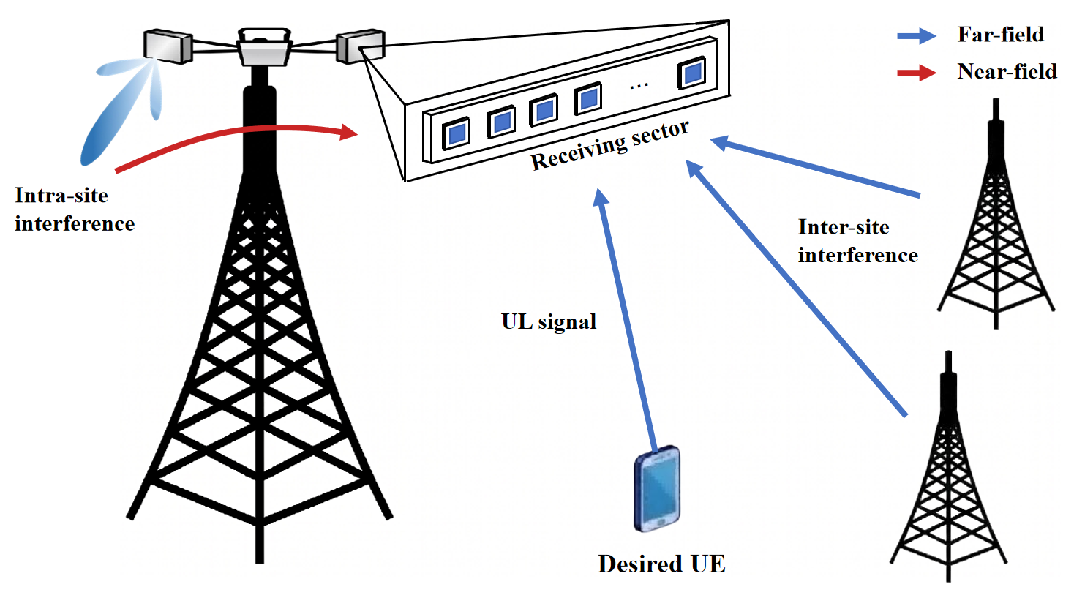}
    \caption{System model for uplink communication under hybrid-field interference.}
    \label{fig:model}
\end{figure}

As illustrated in Fig. \ref{fig:model}, we consider an uplink communication system under hybrid‑field interference.
The hybrid‑field interference consists of near‑field intra‑site interference from co‑sited sectors and far‑field
inter‑site interference from adjacent BSs. The receiving sector is equipped with an $N_{\mathrm r}$‑element uniform linear
array (ULA) and serves a single‑antenna user equipment (UE) over the uplink. 
Let $t$ denote the discrete time index within one coherence
interval. The received signal vector $\mathbf{y}_{\mathrm{BS}}(t) \in \mathbb{C}^{N_{\mathrm{r}} \times 1}$ at the receiving BS sector is modeled as
\begin{equation}
\label{eq:signal_model}
\begin{aligned}
\mathbf{y}_{\mathrm{BS}}(t)= & \underbrace{\mathbf{h}_{\mathrm{UE}} x(t)}_{\text {desired uplink signal }}+\underbrace{\sum_{i=1}^{N_{\mathrm{s}}} \mathbf{h}_i^{\rm {intra }} s_i^{\rm {intra }}(t)}_{\text {intra-site interference }} \\
& +\underbrace{\sum_{k=1}^{L_{\rm {inter }}} \mathbf{h}_k^{\rm {inter }} s_k^{\rm {inter }}(t)}_{\text{inter-site interference }}+\underbrace{\mathbf{n}(t)}_{\text {noise }} ,
\end{aligned}
\end{equation}
where $\mathbf{h}_{\mathrm{UE}}\in\mathbb{C}^{N_{\mathrm r}\times 1}$ denotes the desired uplink channel, $\mathbf{h}^{\mathrm{intra}}_i\in\mathbb{C}^{N_{\mathrm r}\times 1}$ denotes the intra-site interference channel from the $i$-th co-sited  sector, and $\mathbf{h}^{\mathrm{inter}}_k\in\mathbb{C}^{N_{\mathrm r}\times 1}$ denotes the inter-site interference channel from the $k$-th adjacent BS. $N_{\mathrm{s}}$ and $L_{\rm {inter}}$ denote the numbers of intra-site and inter-site interference components, respectively. The symbols $x(t)$, $s_i^{\rm {intra }}(t)$, and $s_k^{\rm {inter }}(t)$ are power-normalized transmitted symbols. The noise vector follows $\mathbf{n}(t)\sim\mathcal{CN}(\mathbf{0},\sigma_n^2\mathbf{I}_{N_{\mathrm r}})$. Other UEs are scheduled on nonoverlapping pilot resources. Given their different propagation scales relative to the receive array, the channel components in \eqref{eq:signal_model} are modeled separately. 

\subsection{Channel Model}
In the considered scenario, the UE is located in the far-field region of the receive array. The desired uplink channel is modeled by a
multipath plane-wave model, given by
\begin{equation}
\begin{aligned}\label{hue}
\mathbf{h}_{\mathrm{UE}} = \sqrt{\frac{N_{\mathrm{r}}}{L}} \sum_{l=1}^{L} \alpha_{l} \mathbf{a}_{N_{\mathrm{r}}}(\theta_l),
\end{aligned}
\end{equation}
where $L$ is the number of propagation paths, and $\alpha_l \sim \mathcal{C N}(0,1 / L)$ and $\theta_l$ denote the complex gain and the angle of arrival (AoA) of the $l$-th path, respectively. Let $d$ denote the antenna spacing, and let the geometric center of the ULA be the phase reference. The physical coordinate of the $n$-th antenna is given by $\delta_n = (n - \frac{N_{\mathrm{r}}-1}{2})d$, $n \in \{0, 1, \dots, N_{\mathrm{r}}-1\}$. The far-field steering vector $\mathbf{a}_{N_{\mathrm{r}}}(\theta) \in \mathbb{C}^{N_{\mathrm{r}} \times 1}$ is expressed as
\begin{equation}
\begin{aligned}
\mathbf{a}_{N_{\mathrm{r}}}(\theta) = \frac{1}{\sqrt{N_{\mathrm{r}}}} \left[ e^{j \frac{2\pi}{\lambda} \delta_0 \sin\theta}, \dots, e^{j \frac{2\pi}{\lambda} \delta_{N_{\mathrm{r}}-1} \sin\theta} \right]^T,
\end{aligned}
\end{equation}
where $\lambda$ is the carrier wavelength. The received power of the desired uplink signal is given by $P_{\mathrm{UE}} = \|\mathbf{h}_{\mathrm{UE}}\|_2^2$.

The intra-site interference originates from co-sited sectors with shorter propagation distances relative to the receive XL-array. Thus, the channel of each intra-site component is modeled by a spherical-wave response, formulated as
\begin{equation}
\begin{aligned}
\mathbf{h}^{\mathrm{intra}}_i = \beta_i \mathbf{b}_{N_{\mathrm{r}}}(\theta_i, r_i),
\end{aligned}
\end{equation}
where $\beta_i$, $\theta_i$, and $r_i$ denote the complex gain, angle, and distance of the interference from the $i$-th co-sited sector, respectively. The magnitude of  \(\beta_i\) is set by the prescribed INR, while its phase is independently randomized. The near-field steering vector $\mathbf{b}_{N_{\mathrm{r}}}(\theta, r)\! \in \!\mathbb{C}^{N_{\mathrm{r}} \times 1}$ is defined as \cite{e1}
\begin{equation}
\begin{aligned}
\mathbf{b}_{N_{\mathrm{r}}}(\theta, r) = \frac{1}{\sqrt{N_{\mathrm{r}}}} \left[ e^{-j \frac{2\pi}{\lambda} (r_n - r)} \right]_{n=0, \dots, N_{\mathrm{r}}-1}^T,
\end{aligned}
\end{equation}
where $r_n = \sqrt{r^2 + \delta_n^2 - 2 r \delta_n \sin\theta}$. The received power from the $i$-th component is $P_{i,\mathrm{intra}} = |\beta_i|^2$, and the aggregate intra-site interference power is $P_{\mathrm{intra}} = \sum_{i=1}^{N_{\mathrm{s}}} |\beta_i|^2$. Due to the short propagation distance, the intra-site interference is typically strong and varies slowly over the channel estimation interval.

For the inter-site interference, the channel $\mathbf{h}^{\mathrm{inter}}_k$ from the $k$-th adjacent BS is modeled as 
$
\mathbf{h}_k^{\rm {inter }}=\gamma_k \mathbf{a}_{N_{\mathrm r}}\left(\theta_k^{\rm {inter }}\right),
$
where $\gamma_k$ and $\theta_k^{\rm {inter }}$ denote the corresponding complex gain and AoA, respectively. Although the inter-site interference is generally weaker than intra-site interference because of longer propagation distances, its residual contribution can still bias the desired uplink channel estimate.

\subsection{Limitations of Conventional Channel Estimation}
In practice, traditional channel estimation correlates the received signal with the desired uplink
pilot. We first consider a conventional LS estimator, given by
\begin{equation}
\begin{aligned}\label{eq:ls_estimation}
&\hat{\mathbf{h}}_{\mathrm{LS}}=\frac{1}{\tau} \sum_{t=1}^\tau \mathbf{y}_{\mathrm{BS}}(t) x^{\star}(t)=\mathbf{h}_{\mathrm{UE}}+\mathbf{i}_{\mathrm{res}}+\tilde{\mathbf{n}},\\
\end{aligned}
\end{equation}
where the residual interference term $\mathbf{i}_{\mathrm{res}}$ is given by
\begin{equation}
\begin{aligned}
&\mathbf{i}_{\rm {res }}=\sum_{i=1}^{N_{\mathrm s}} \rho_i^{\rm {intra }} \mathbf{h}_i^{\rm {intra }}+\sum_{k=1}^{L_{\rm {inter }}} \rho_k^{\rm {inter }} \mathbf{h}_k^{\rm {inter }} .
\end{aligned}
\end{equation}
Here, $\tau$ denotes the pilot length, $\tilde{\mathbf{n}}\sim
\mathcal{CN}
\left(
\mathbf{0},
\frac{\sigma_n^2}{\tau}\mathbf{I}_{N_{\mathrm r}}
\right)$ denotes the effective noise, the coefficients
$\rho_i^{\rm intra}
\!=\!
\frac{1}{\tau}
\sum_{t=1}^{\tau}
s_i^{\rm intra}(t)x^{\star}(t),
$ and $\rho_k^{\rm inter}
\!=\!\frac{1}{\tau}
\sum_{t=1}^{\tau}
s_k^{\rm inter}(t)x^{\star}(t)$
denote the finite-length correlations between the desired pilot and the interference symbols. Since the interference signals are not coordinated with the desired uplink pilot, these correlations do not vanish for a finite pilot length. As a result, the LS error contains a residual hybrid-field interference term.

To reduce this residual interference, a covariance-based linear MMSE estimator can  be applied as
\begin{equation}
\begin{aligned}
\label{eq:lmmse_estimator}
&\hat{\mathbf h}_{\rm LMMSE}
=
\mathbf W\hat{\mathbf h}_{\rm LS},\\
&\mathbf W
=
\mathbf R_{\rm UE}
\left(
\mathbf R_{\rm UE}
+
\mathbf R_{\rm int}
+
\sigma_{\tilde n}^2\mathbf I_{N_{\rm r}}
\right)^{-1},
\end{aligned}
\end{equation}
where
\(\mathbf R_{\rm UE}=\mathbb E[\mathbf h_{\rm UE}\mathbf h_{\rm UE}^H]\) and 
\(\mathbf R_{\rm int}=\mathbb E[\mathbf i_{\rm res}\mathbf i_{\rm res}^H]\). 
\(\sigma_{\tilde n}^{2}=\sigma_n^{2}/\tau\) is the effective noise variance. Assuming that \(\mathbf h_{\rm UE}\), \(\mathbf i_{\rm res}\), and \(\tilde{\mathbf n}\) are mutually uncorrelated, the MSE is
\begin{equation}
\begin{aligned}
{\rm MSE}_{\rm LMMSE}
=&\operatorname{tr}\!\left((\mathbf W-\mathbf I_{N_{\mathrm r}})
\mathbf R_{\rm UE}
(\mathbf W-\mathbf I_{N_{\mathrm r}})^H\right)  \\
&+\operatorname{tr}\!\left(\mathbf W\mathbf R_{\rm int}\mathbf W^H\right)
+\sigma_{\tilde n}^2\operatorname{tr}\!\left(\mathbf W\mathbf W^H\right).
\end{aligned}
\end{equation}

This decomposition shows that the error is not merely noise-limited. When the residual hybrid-field interference is strong, the second term is governed by \(\mathbf R_{\rm int}\).
The above analysis shows that conventional pilot-based estimators remain structurally limited under strong hybrid-field interference, motivating interference suppression before desired uplink channel estimation.

\section{Hybrid-Field Interference Suppression}\label{sec3}

In this section, we introduce a dictionary-based framework for
hybrid-field interference suppression. The received observations are
represented over a hybrid-field dictionary to reconstruct dominant
interference components, whose subspace is then nulled before
desired uplink channel reconstruction.

\subsection{Dictionary-Based Sparse Reconstruction for Interference Suppression}

To suppress hybrid‑field interference before channel estimation, the received observations are represented over a hybrid-field dictionary\cite{zeng2025hybrid}
\begin{equation}
\label{eq:mf_dictionary}
\mathbf D
\triangleq
[
\mathbf D_{\mathrm{NF}}, 
\mathbf F_{\rm A}
],
\end{equation}
where \(\mathbf D_{\mathrm{NF}}\) contains near-field atoms indexed by angle and distance for intra-site interference, and \(\mathbf F_{\rm A}\) contains angular atoms for inter-site interference and the desired uplink channel.
Specifically, the dictionary $\mathbf F_{\rm A}\in\mathbb{C}^{N_{\mathrm r}\times G_{\mathrm f}}$ is constructed by sampling the angular interval $[-\pi/2,\pi/2]$ with $G_{\mathrm f}$ far-field steering vectors, given by
\begin{equation}
\label{eq:FA_matrix}
\mathbf F_{\rm A}
=
[
\mathbf a_{N_{\mathrm r}}(\theta_1),
\mathbf a_{N_{\mathrm r}}(\theta_2),
\ldots,
\mathbf a_{N_{\mathrm r}}(\theta_{G_{\mathrm f}})
].
\end{equation}
The near-field dictionary $\mathbf D_{\mathrm{NF}}\in\mathbb{C}^{N_{\mathrm r}\times G_{\mathrm n}}$ is constructed by discretizing the polar-domain coordinate $\mathbf q=[\theta,r]^T$:
\begin{equation}
\label{eq:cs_dictionary}
\mathbf D_{\mathrm{NF}}
=
[
\mathbf b_{N_{\mathrm r}}(\mathbf q_1),
\ldots,
\mathbf b_{N_{\mathrm r}}(\mathbf q_{G_{\mathrm n}})
],
\end{equation}
where \(\mathbf b_{N_{\mathrm r}}(\mathbf q)\) denotes the
near-field steering vector parameterized by \(\mathbf q\), and
\(G_n=G_\theta G_r\) is the number of near-field grid points,
with \(G_\theta\) angle samples and \(G_r\) distance samples.

Let
$\mathbf Y
=
[\mathbf y_{\mathrm{BS}}(1),\ldots,\mathbf y_{\mathrm{BS}}(\tau)]
\in\mathbb{C}^{N_{\mathrm r}\times\tau}$
denote the received observation matrix during the pilot interval, and let $\mathbf d_i$ be the $i$-th atom of $\mathbf D$. The hybrid-field OMP procedure initializes $\mathbf E^{(0)}=\mathbf Y$ and selects the dominant hybrid-field interference atom at the $p$-th iteration as
\begin{equation}
\label{eq:mf_omp_selection}
i_p^{\star}
=
\arg\max_i
\left\|
\mathbf d_i^H
\mathbf E^{(p-1)}
\right\|_2^2 .
\end{equation}
The selected atoms are collected as
$\hat{\mathbf A}_{\mathrm{int}}^{(p)}
=
[\mathbf d_{i_1^{\star}},\ldots,\mathbf d_{i_p^{\star}}]$,
and the residual is updated by
\begin{equation}
\label{eq:int_residual_update}
\mathbf E^{(p)}
=
\mathbf Y
-
\hat{\mathbf A}_{\mathrm{int}}^{(p)}
\left(
\hat{\mathbf A}_{\mathrm{int}}^{(p)}
\right)^{\dagger}
\mathbf Y .
\end{equation}
After \(K_{\mathrm{int}}\) iterations, we set
$\hat{\mathbf A}_{\mathrm{int}}=\hat{\mathbf A}_{\mathrm{int}}^{(K_{\mathrm{int}})}$, where
\(K_{\mathrm{int}}\) denotes the interference order selected by
the reconstruction procedure.
The reconstructed subspace is nulled by
\begin{equation}
\label{eq:apn_projection}
\mathbf P_{\mathrm{int}}^{\perp}
=
\mathbf I_{N_{\mathrm r}}
-
\hat{\mathbf A}_{\mathrm{int}}
\hat{\mathbf A}_{\mathrm{int}}^{\dagger},
\quad
\mathbf y_{\mathrm p}
=
\mathbf P_{\mathrm{int}}^{\perp}
\hat{\mathbf h}_{\mathrm{LS}},
\quad
\mathbf U_{\mathrm p}
=
\mathbf P_{\mathrm{int}}^{\perp}
\mathbf F_{\rm A}, 
\end{equation}
where $\mathbf y_{\mathrm p}$ is the interference-nulled LS estimate, and $\mathbf U_{\mathrm p}$ is the projected far-field dictionary.

Let $\mathbf u_{\mathrm p,m}$ denote the $m$-th atom of $\mathbf U_{\mathrm p}$. With $\mathbf r_{\mathrm{UE}}^{(0)}=\mathbf y_{\mathrm p}$, the $l$-th UE path is selected by the normalized correlation
\begin{equation}
\label{eq:ue_omp_selection}
m_l^{\star}
=
\arg\max_{m:\|\mathbf u_{\mathrm p,m}\|_2\neq 0}
\frac{
\left|
\mathbf u_{\mathrm p,m}^H
\mathbf r_{\mathrm{UE}}^{(l-1)}
\right|
}{
\|\mathbf u_{\mathrm p,m}\|_2
}.
\end{equation}
After selecting the first $l$ atoms, let
$\hat{\mathbf U}_{\mathrm p}^{(l)}
=
[\mathbf u_{\mathrm p,m_1^{\star}},\ldots,\mathbf u_{\mathrm p,m_l^{\star}}]$.
The residual is updated as
\begin{equation}
\label{eq:ue_residual_update}
\mathbf r_{\mathrm{UE}}^{(l)}
=
\mathbf y_{\mathrm p}
-
\hat{\mathbf U}_{\mathrm p}^{(l)}
\left(
\hat{\mathbf U}_{\mathrm p}^{(l)}
\right)^{\dagger}
\mathbf y_{\mathrm p}.
\end{equation}

After $L$ UE paths are selected, the support set is recovered as 
$\hat{\mathcal S}_{\mathrm{UE}}
=
\{m_1^{\star},\ldots,m_{ L}^{\star}\}$,
and the corresponding far-field basis is
$\hat{\mathbf A}_{\mathrm{UE}}
=
\mathbf F_{\rm A}(:,\hat{\mathcal S}_{\mathrm{UE}})$.
Under the interference‑nulling projection, the UE path gains and the desired uplink channel are obtained as
\begin{equation}
\label{eq:ue_reconstruction}
\hat{\mathbf g}_{\mathrm{UE}}
=
\hat{\mathbf A}_{\mathrm{UE}}^{\dagger}
\mathbf y_{\mathrm p},
\quad
\hat{\mathbf h}_{\mathrm{UE}}
=
\hat{\mathbf A}_{\mathrm{UE}}
\hat{\mathbf g}_{\mathrm{UE}} .
\end{equation}

\begin{figure}[t]
    \centering
\includegraphics[width=1\linewidth]{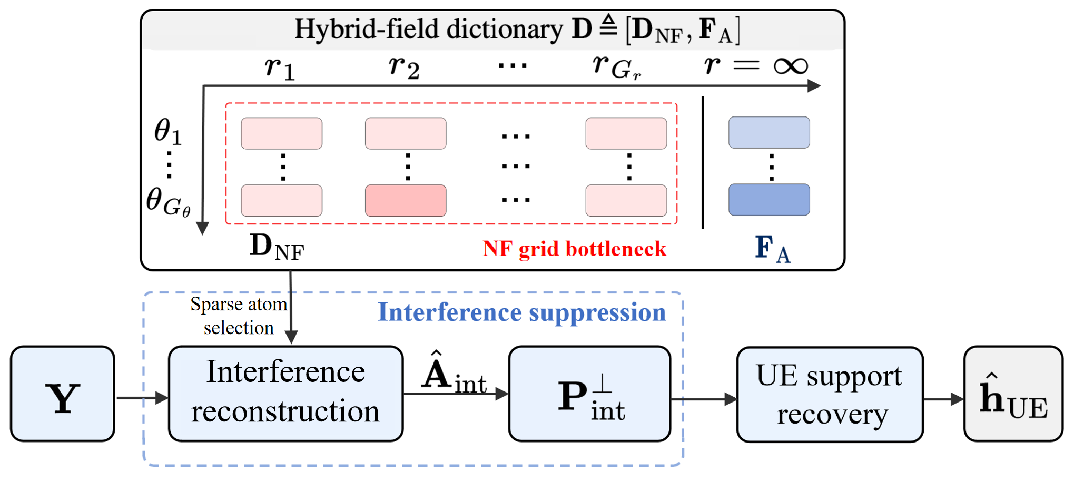}
   \caption{Dictionary-based sparse reconstruction framework for hybrid-field interference suppression. Dominant interference atoms are selected from the unified hybrid-field dictionary \(\mathbf D\). The highlighted near-field angle--distance matching is the bottleneck analyzed in Section III-B.}
    \label{fig:framework}
\end{figure}

Although hybrid-field interference reconstruction is performed over the unified dictionary \(\mathbf D\), the main bottleneck is the exhaustive atom matching over the
finely discretized angle–distance grid used to construct \(\mathbf D_{\rm NF}\). The
highlighted 2D fine-grid search in Fig.~\ref{fig:framework} leads to the complexity and coherence limitations analyzed next.

\subsection{Limitations of Dictionary‑Based Interference Suppression}

\textbf{1) High Complexity from Fine-Grid Search: }
Let $\mathbf{q}_i^{\star}=\left[\theta_i, r_i\right]^T$ denote the true angle--distance coordinate and let $\mathbf{q}_g=\mathbf{q}_i^{\star}+\Delta \mathbf{q}$ be its nearest grid point in $\mathbf{D}_{\mathrm{NF}}$. The grid mismatch is measured as a second-order expansion of the near-field manifold, given by

\begin{equation}
1-\left|\mathbf{b}_{N_{\mathrm r}}^H\left(\mathbf{q}_i^{\star}\right) \mathbf{b}_{N_{\mathrm r}}\left(\mathbf{q}_g\right)\right|^2 \approx \Delta \mathbf{q}^T \mathbf{J}\left(\mathbf{q}_i^{\star}\right) \Delta \mathbf{q},
\end{equation}
where $\mathbf J(\mathbf q_i^{\star})\in\mathbb{R}^{2\times 2}$ is the local metric tensor of the near-field manifold \cite{e5}. The residual interference power after nulling is approximately
$
P_{\rm {\rm{res}}, i} \approx P_{i, \rm { intra }} \Delta \mathbf{q}^T \mathbf{J}\left(\mathbf{q}_i^{\star}\right) \Delta \mathbf{q} .
$
To keep this residual power below the effective noise floor, i.e., $P_{\rm {\rm{res}}, i} \leq \xi \sigma_{\tilde{n}}^2$, the grid error must satisfy
\begin{equation}
\label{eq:grid_bound}
\Delta\mathbf q^T
\mathbf J(\mathbf q_i^{\star})
\Delta\mathbf q
\le
\xi
\frac{\sigma_{\tilde n}^2}{P_{i,\mathrm{intra}}}
=
\xi\mathrm{INR}_i^{-1},
\end{equation}
where $\mathrm{INR}_i\triangleq P_{i,\mathrm{intra}}/\sigma_{\tilde n}^2$ and $\xi \geq 1$ is a tolerance factor.
For large \(\mathrm{INR}_i\), \eqref{eq:grid_bound} requires a fine near‑field angle‑distance grid, which increases \(G_{\mathrm n}=G_\theta
G_r\). Since the hybrid‑field dictionary has size \(G_{\rm H}=G_{\mathrm n}+G_{\mathrm f}\), the per‑iteration correlation cost of OMP over
\(\mathbf D\) scales as
$\mathcal O(N_{\mathrm r}\tau G_{\rm H})$.
Therefore, the fine grid required for strong interference suppression increases the computational complexity.

\textbf{2) Sparse Recovery Degradation from Dictionary Coherence:}
Furthermore, refining the near-field angle--distance grid increases the correlation between adjacent dictionary atoms~\cite{chi2011basis,tropp2004greed}. The dictionary coherence of $\mathbf D_{\mathrm{NF}}$ is defined as
\begin{equation}
\mu(\mathbf D_{\mathrm{NF}})
\triangleq
\max_{u\neq v}
\left|
\mathbf d_u^H\mathbf d_v
\right|,
\end{equation}
where $\mathbf d_u$ and $\mathbf d_v$ are atoms of $\mathbf D_{\mathrm{NF}}$. For two adjacent grid points separated by $\Delta\mathbf q$, the local approximation gives
\begin{equation}
\label{eq:mutual_coherence}
\mu(\mathbf D_{\mathrm{NF}})
\approx
1
-
\frac{1}{2}
\Delta\mathbf q^T
\mathbf J(\mathbf q_i^{\star})
\Delta\mathbf q
\ge
1-\frac{\xi}{2\mathrm{INR}_i}.
\end{equation}
Thus, the fine grid required to reduce residual interference can drive the dictionary coherence close to 1 in high‑INR regimes, which degrades sparse recovery.
\begin{figure}
    \centering
\includegraphics[width=1\linewidth]{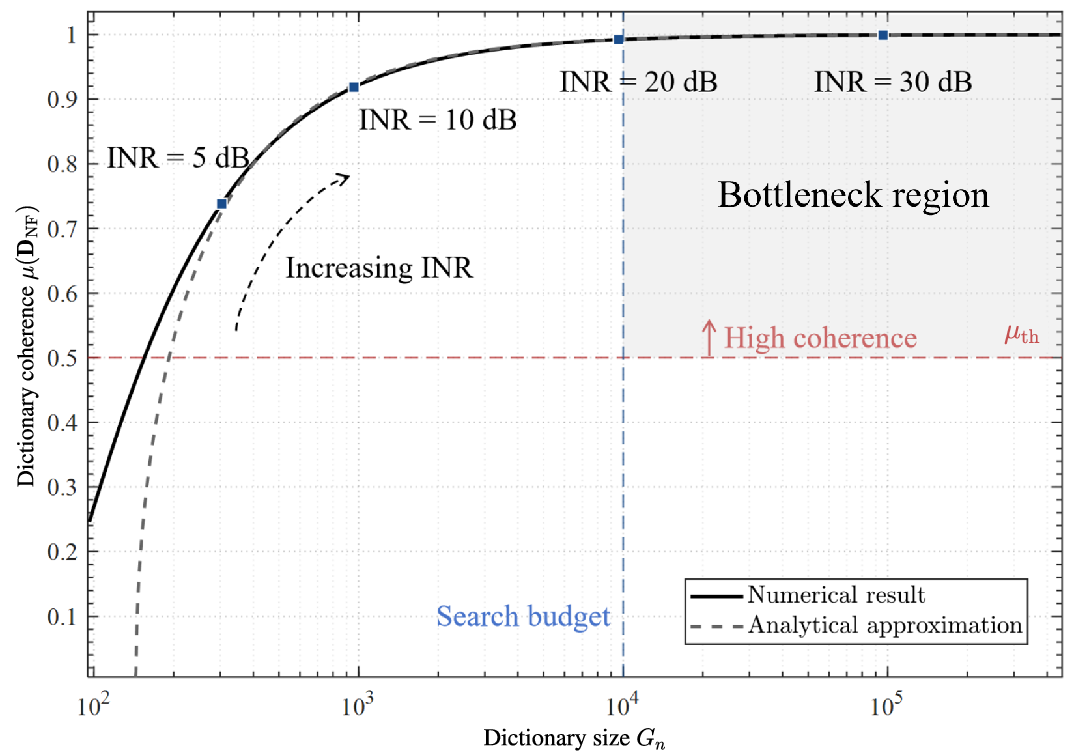}
    \caption{Bottleneck region of dictionary-based near-field angle--distance matching. Increasing intra-site INR requires a finer near-field grid to satisfy the residual-interference bound, while grid refinement increases \(\mu(\mathbf D_{\rm NF})\) and may exceed the search budget.}
    \label{fig:cplx}
\end{figure}

\textbf{Remark 1 (Bottleneck Region):}
As illustrated in Fig.~\ref{fig:cplx}, increasing INR requires a finer angle--distance grid to satisfy the interference bound in \eqref{eq:grid_bound}. For illustration, the plot uses a coherence threshold of
\(\mu_{\mathrm{th}}=0.5\) and a search budget of
\(10^4\) near-field atoms. However, this refinement increases both the near‑field dictionary size \(G_{\mathrm n}\) and the coherence \(\mu(\mathbf D_{\rm NF})\) in \eqref{eq:mutual_coherence}. When the required \(G_{\mathrm n}\)
exceeds the available search budget and \(\mu(\mathbf D_{\rm NF})\) approaches a prescribed threshold \(\mu_{\rm th}\),
dictionary‑based search becomes impractical.

These limitations motivate replacing dense near-field
angle--distance matching with interference-parameter estimation.

\section{Low-Complexity Parameter Estimation for Hybrid-Field Interference Reconstruction}\label{sec4}

To address the near-field search bottleneck discussed in Section III, this section develops a low-complexity interference-parameter estimation method. First, the received spatial signal is mapped onto a 1D far-field angular spectrum, where the dominant near-field interference forms a structured diffusion envelope. The interference angle is obtained from envelope symmetry, and the interference distance is estimated through monotone power--curvature inversion. The monotone inversion range is further translated into a physical critical-distance bound to characterize the feasible operating region.

\subsection{Spectral Dominance and Angular Diffusion Modeling}

We first characterize the angular power spectrum (APS) of the received signal and establish the dominance of near-field intra-site interference under the high-INR condition.
The sample covariance matrix (SCM) over $\tau$ snapshots is
\begin{equation}
\hat{\mathbf R}_{\mathbf y}
=
\frac{1}{\tau}
\sum_{t=1}^{\tau}
\mathbf y_{\mathrm{BS}}(t)
\mathbf y_{\mathrm{BS}}^H(t).
\end{equation}
The APS evaluated at the scanning angle $\theta_m$ is
\begin{equation}
\label{eq:aps_expansion}
\begin{aligned}
\mathcal P(\theta_m)
=&~
\mathbf a_{N_{\mathrm r}}^H(\theta_m)
\hat{\mathbf R}_{\mathbf y}
\mathbf a_{N_{\mathrm r}}(\theta_m)  \\
=&~
\mathcal P_{\mathrm{intra}}(\theta_m)
+
\mathcal P_{\mathrm{inter}}(\theta_m)
+
\mathcal P_{\mathrm{UE}}(\theta_m)\\
&+
\mathcal P_{\mathrm n}(\theta_m)
+
\mathcal P_{\mathrm{cross}}(\theta_m),
\end{aligned}
\end{equation}
where
$\mathcal P_{\chi}(\theta_m)
\triangleq
\mathbf a_{N_{\mathrm r}}^H(\theta_m)
\hat{\mathbf R}_{\chi}
\mathbf a_{N_{\mathrm r}}(\theta_m)$
denotes the projected angular power of component $\chi\in\{\mathrm{intra},\mathrm{inter},\mathrm{UE},\mathrm n\}$.
The term $\mathcal P_{\mathrm{cross}}(\theta_m)$ collects the finite-snapshot cross-covariance terms among different components. This decomposition motivates the following key result.

\textbf{Proposition 1 (Spectral Dominance):}
In the strong intra-site interference regime, i.e.,
$P_{\mathrm{intra}}\!\gg\! P_{\mathbf z}$ with $P_{\mathbf z}$ denoting the aggregate power of the non-dominant components, the composite APS within the main diffusion envelope of the intra-site interference is dominated by $\mathcal P_{\mathrm{intra}}(\theta_m)$.

\begin{IEEEproof}
Let $\mathbf z(t)$ denote the aggregate non-dominant component, including the inter-site interference, the desired uplink signal, and the noise. Its SCM is defined as
$\hat{\mathbf R}_{\mathbf z}
\triangleq
\frac{1}{\tau}
\sum_{t=1}^{\tau}
\mathbf z(t)\mathbf z^H(t)$,
with
$P_{\mathbf z}\triangleq\mathrm{tr}(\hat{\mathbf R}_{\mathbf z})$.
Since $\hat{\mathbf R}_{\mathbf z}$ is positive semidefinite, the Rayleigh quotient gives
\begin{equation}
\label{eq:rayleigh_bound}
\mathcal P_{\mathbf z}(\theta_m)
=
\mathbf a_{N_{\mathrm r}}^H(\theta_m)
\hat{\mathbf R}_{\mathbf z}
\mathbf a_{N_{\mathrm r}}(\theta_m)
\le
\lambda_{\max}(\hat{\mathbf R}_{\mathbf z})
\le
P_{\mathbf z}.
\end{equation}The first inequality becomes equality when
\(\mathbf a_{N_{\mathrm r}}(\theta_m)\) lies in the principal
eigenspace of \(\hat{\mathbf R}_{\mathbf z}\), whereas the second
becomes equality when $\mathrm{rank}(\hat{\mathbf R}_{\mathbf z}) \le 1$. 
Within the main diffusion envelope of the intra-site interference, assume
$\mathcal P_{\mathrm{intra}}(\theta_m)$ remains on the order of
$P_{\mathrm{intra}}$. Under $P_{\mathrm{intra}}\gg P_{\mathbf z}$, it follows that
$\mathcal P_{\mathrm{intra}}(\theta_m)\gg \mathcal P_{\mathbf z}(\theta_m)$. By the Cauchy-Schwarz inequality, the cross term satisfies
\begin{equation}
\label{eq:cross_term_bound}
|
\mathcal P_{\mathrm{cross}}(\theta_m)
|
\le
2
\sqrt{
\mathcal P_{\mathrm{intra}}(\theta_m)
\mathcal P_{\mathbf z}(\theta_m)
}
\le
2
\sqrt{
\mathcal P_{\mathrm{intra}}(\theta_m)
P_{\mathbf z}
}.
\end{equation}
Define
$\epsilon(\theta_m)
\triangleq
\sqrt{
\mathcal P_{\mathbf z}(\theta_m)/
\mathcal P_{\mathrm{intra}}(\theta_m)
}$.
Then $\epsilon(\theta_m)\ll 1$ holds within the main diffusion envelope, and normalizing \eqref{eq:aps_expansion} by
$\mathcal P_{\mathrm{intra}}(\theta_m)$ yields
\begin{equation}
\label{eq:squeeze_bound}
(1-\epsilon(\theta_m))^2
\le
\frac{\mathcal P(\theta_m)}
{\mathcal P_{\mathrm{intra}}(\theta_m)}
\le
(1+\epsilon(\theta_m))^2 .
\end{equation}
Therefore,
$\mathcal P(\theta_m)
=
\mathcal P_{\mathrm{intra}}(\theta_m)
(1+\mathcal O(\epsilon(\theta_m)))$,
which establishes the dominance of the intra-site term.
\end{IEEEproof}

Proposition~1 shows that the composite APS is dominated by the near-field intra-site interference. Under the considered high-INR
regime, the near-field interference is therefore estimated and nulled first,
reducing the residual dynamic range before far-field interference
reconstruction. We then characterize its angular diffusion.

For the $i$-th intra-site interference source, define the normalized far-field angular response as
\begin{equation}
\label{eq:spatial_response_def}
\begin{aligned}
\Gamma_{N_{\mathrm r}}(\theta_m)
&=
\left|
\mathbf a_{N_{\mathrm r}}^H(\theta_m)
\mathbf b_{N_{\mathrm r}}(\theta_i,r_i)
\right|^2  \\
&=
\left|
\frac{1}{N_{\mathrm r}}
\sum_{n=0}^{N_{\mathrm r}-1}
\exp
\left[
-j\frac{2\pi}{\lambda}
\left(
\delta_n\sin\theta_m
+
r_n-r_i
\right)
\right]
\right|^2 .
\end{aligned}
\end{equation}
Using the second-order expansion \cite{you114}
\begin{equation}
r_n-r_i
\approx
-\delta_n\sin\theta_i
+
\frac{\delta_n^2\cos^2\theta_i}{2r_i},
\end{equation}
and defining the effective aperture
$D_{\mathrm a}\triangleq (N_{\mathrm r}-1)d$ and the normalized coordinate
$x_n\triangleq 2\delta_n/D_{\mathrm a}\in[-1,1]$,
\eqref{eq:spatial_response_def} becomes
\begin{equation}
\label{eq:quadratic_phase_response}
\Gamma_{N_{\mathrm r}}(\theta_m)
\approx
\left|
\frac{1}{N_{\mathrm r}}
\sum_{n=0}^{N_{\mathrm r}-1}
\exp
\left[
j\frac{\pi}{2}
\left(
M x_n
-
K x_n^2
\right)
\right]
\right|^2 ,
\end{equation}
where
\begin{equation}
\label{eq:MK_def}
M
\triangleq
-\frac{2D_{\mathrm a}}{\lambda}
(\sin\theta_m-\sin\theta_i),
\quad
K
\triangleq
\frac{D_{\mathrm a}^2\cos^2\theta_i}{2\lambda r_i}.
\end{equation}
Here, $M$ represents the angular mismatch in the linear phase term, while $K$ represents the distance-dependent wavefront curvature. The resulting quadratic-phase sum determines the angular diffusion envelope, which is given below.

\textbf{Theorem 1 (Fresnel Approximation):}
Under the radiating near-field condition, for a sufficiently large ULA, the normalized discrete response in \eqref{eq:quadratic_phase_response} is approximated by
\begin{equation}
\label{eq:fresnel_response}
\Gamma(M,K)
\approx
\frac{1}{4K}
\left[
\Delta\mathcal C^2(M,K)
+
\Delta\mathcal S^2(M,K)
\right],
\quad K>0,
\end{equation}
where $\mathcal{C}(z) = \int_0^z \cos(\frac{\pi}{2}u^2)du$ and $\mathcal{S}(z) = \int_0^z \sin(\frac{\pi}{2}u^2)du$ denote the Fresnel integrals, and $\Delta\mathcal C(M,K)=\mathcal C(t_2)-\mathcal C(t_1)$ and
$\Delta\mathcal S(M,K)=\mathcal S(t_2)-\mathcal S(t_1)$.
The integration limits are
\begin{equation}
\label{eq:integration_limits}
t_1
=
-\sqrt K-\frac{M}{2\sqrt K},
\quad
t_2
=
\sqrt K-\frac{M}{2\sqrt K}.
\end{equation}

\begin{IEEEproof}
For large $N_{\mathrm r}$, the normalized coordinates $x_n$ form a dense uniform sampling grid over $[-1,1]$. Hence, the discrete average in \eqref{eq:quadratic_phase_response} can be approximated by the corresponding Riemann integral, given by
\begin{equation}
\label{eq:riemann}
\frac{1}{N_{\mathrm r}}
\sum_{n=0}^{N_{\mathrm r}-1}
e^{j\frac{\pi}{2}(Mx_n-Kx_n^2)}
\approx
\frac{1}{2}
\int_{-1}^{1}
e^{j\frac{\pi}{2}(Mx-Kx^2)}
dx.
\end{equation}
Completing the square  for the quadratic phase argument yields
$
-Kx^2+Mx
=
-K
\left(
x-\frac{M}{2K}
\right)^2
+
\frac{M^2}{4K}.
$
With the substitution
$t=\sqrt K(x-M/(2K))$ and $dx=K^{-1/2}dt$, the integral in \eqref{eq:riemann} can be rewritten as
\begin{equation}
\label{eq:fresnel_transform}
\begin{aligned}
&\frac{e^{j \frac{\pi M^2}{8K}}}{2\sqrt{K}} \int_{t_1}^{t_2} e^{-j \frac{\pi}{2} t^2} dt \\ 
&= \frac{e^{j \frac{\pi M^2}{8K}}}{2\sqrt{K}} \left[ \int_{t_1}^{t_2} \cos\left(\frac{\pi}{2}t^2\right) dt - j \int_{t_1}^{t_2} \sin\left(\frac{\pi}{2}t^2\right) dt \right] \\ 
&= \frac{e^{j \frac{\pi M^2}{8K}}}{2\sqrt{K}} \Big[ \big( \mathcal{C}(t_2) - \mathcal{C}(t_1) \big) - j \big( \mathcal{S}(t_2) - \mathcal{S}(t_1) \big) \Big]. 
\end{aligned}
\end{equation}
Taking the squared magnitude removes the unit-modulus phase term and yields \eqref{eq:fresnel_response}.
\end{IEEEproof}

Theorem 1 shows that the distance-dependent curvature $K$ prevents a near-field component from being represented by a single far-field angular atom. Instead, its energy follows a structured diffusion profile in the 1D angular spectrum. This property provides the basis for analytical estimation of the angle and distance of near-field components.

\subsection{Interference Parameter Estimation from Angular Diffusion}

Motivated by Proposition 1, the dominant intra-site interference is extracted before the residual inter-site interference. Unlike far-field inter-site interference, which is sparse in the angular domain, near-field intra-site interference produces a diffused angular-spectrum envelope. This subsection focuses on estimating its spatial parameters, i.e., the angle $\theta_i$ and distance $r_i$. The inter-site components are estimated after the dominant intra-site interference subspace is nulled.

\subsubsection{Angle Estimation from Envelope Symmetry}

Although near-field diffusion prevents a sparse angular representation, its envelope admits a sine-domain symmetry.

\textbf{Lemma 1 (Symmetry of Near-Field Angular Diffusion):}
For a given curvature factor $K$, the power response exhibits symmetry with respect to the linear angle mismatch $M$, i.e., $\Gamma(-M, K)=\Gamma(M, K)$. Consequently, if two scanning angles $\theta_{+}$ and $\theta_{-}$ denote paired equal-power points on the two sides of the main diffusion envelope, the interference angle $\theta_i$ satisfies
\begin{equation}
\label{eq:sine_domain_symmetry}
\theta_i
=
\arcsin
\left(
\frac{\sin\theta_{+}+\sin\theta_{-}}{2}
\right).
\end{equation}

\begin{IEEEproof}
From \eqref{eq:riemann}, the  response is determined by
\begin{equation}
I(M,K)
=
\frac{1}{2}
\int_{-1}^{1}
e^{j\frac{\pi}{2}(Mx-Kx^2)}
dx .
\end{equation}
For $-M$, applying the substitution $x=-u$ gives
\begin{equation}
\begin{aligned}
I(-M,K)
&=
\frac{1}{2}
\int_{-1}^{1}
e^{j\frac{\pi}{2}(-Mx-Kx^2)}
dx \\
&=
\frac{1}{2}
\int_{-1}^{1}
e^{j\frac{\pi}{2}(Mu-Ku^2)}
du
=
I(M, K).
\end{aligned}
\end{equation}
Therefore, $\Gamma(-M,K)=|I(-M,K)|^2=|I(M,K)|^2=\Gamma(M,K)$.
For two paired equal-power points on opposite sides of the main envelope, the corresponding mismatch parameters satisfy $M_1=-M_2$. Using
$M=-\frac{2D_{\mathrm a}}{\lambda}(\sin\theta_m-\sin\theta_i)$,
we obtain
\begin{equation}
\sin\theta_i
=
\frac{\sin\theta_{+}+\sin\theta_{-}}{2},
\end{equation}
which yields \eqref{eq:sine_domain_symmetry}.
\end{IEEEproof}

The symmetry in Lemma 1 is exact for the second-order quadratic-phase model in
\eqref{eq:quadratic_phase_response}, but is generally approximate for the exact spherical-wave response in \eqref {eq:spatial_response_def}.
Higher-order phase terms omitted by the Fresnel approximation can introduce envelope asymmetry \cite{wu2024dft}. Therefore, Lemma 1 provides a low-complexity approximate angle estimate for intra-site interference.

\subsubsection{Distance Inversion via Monotone Mapping}

After the interference angle is obtained, setting $\theta_m=\theta_i$ eliminates the linear mismatch term, i.e., $M=0$. The two-dimensional response $\Gamma(M, K)$ then reduces to a one-dimensional function of the curvature factor $K$. From \eqref{eq:integration_limits}, we have $t_2=-t_1=\sqrt K$, and hence
\begin{equation}
\label{eq:peak_power_response}
\Gamma(0,K)
=
\frac{
\mathcal C^2(\sqrt K)
+
\mathcal S^2(\sqrt K)
}{K}
\triangleq
\mathcal F(K).
\end{equation}
The function \(\mathcal F(K)\) maps the curvature factor to the normalized peak power at the interference angle.

\textbf{Lemma 2 (Monotone Power--Curvature Mapping):}
Let
\begin{equation}
\label{eq:valid_domain}
\mathcal D_{\mathrm{inv}}
=
\left\{
K
\,\middle|\, K\in(0,K_{\max}],\ 
\frac{d\mathcal F(K)}{dK}<0
\right\}
\end{equation}
denote a monotone inversion interval, where $K_{\max}$ is chosen before the first stationary point of $\mathcal F(K)$. Then $\mathcal F:\mathcal D_{\mathrm{inv}}\rightarrow\mathcal R_{\mathrm{image}}$, where $\mathcal R_{\mathrm{image}}$ is the range of $\mathcal F$, is bijective and admits a smooth inverse. Therefore, the curvature can be recovered as
\begin{equation}
\label{eq:curvature_inverse}
K
=
\mathcal F^{-1}(P_{\mathrm{obs}}),
\end{equation}
where $P_{\mathrm{obs}}\!\in\!\mathcal R_{\mathrm{image}}$ denotes the APS peak at \(\hat\theta_i\) normalized by the corresponding diffusion-lobe area.

\begin{IEEEproof}
Let $z=\sqrt K$. As $K\to 0$, the Fresnel integrals satisfy
$\mathcal C(z)=z+\mathcal O(z^5)$ and
$\mathcal S(z)=\frac{\pi}{6}z^3+\mathcal O(z^7)$.
Thus,
$\mathcal C^2(\sqrt K)+\mathcal S^2(\sqrt K)=K+\mathcal O(K^3)$, and
\begin{equation}
\label{eq:limit_proof}
\lim_{K\to 0}\mathcal F(K)
=
\lim_{K\to 0}
\frac{K+\mathcal O(K^3)}{K}
=
1.
\end{equation}
This corresponds to coherent phase integration in the plane-wave limit.

For $K>0$, both Fresnel integrals are smooth, and therefore $\mathcal F(K)$ is smooth on $\mathcal D_{\mathrm{inv}}$. By construction, $\mathcal F'(K)<0$ on $\mathcal D_{\mathrm{inv}}$, so $\mathcal F(K)$ is strictly monotone and globally injective on this interval. Since $\mathcal F'(K)\neq0$, the inverse function theorem guarantees a smooth inverse on $\mathcal R_{\mathrm{image}}$.
\end{IEEEproof}

Since the monotone mapping $\mathcal F(K)$ does not admit a convenient closed-form inverse, estimating $K$ iteratively for each envelope introduces unnecessary overhead. We therefore approximate the inverse mapping using a precomputed PCHIP \cite{e8}. Specifically, \(S\) curvature samples
\(K_s\in\mathcal D_{\mathrm{inv}}\), \(s=1,\ldots,S\),
are generated, and their normalized responses
\(P_s=\mathcal F(K_s)\) are evaluated. After sorting the inverse pairs by $P_s$, we define

\begin{equation}
\mathcal{T} \triangleq \{ (P_s, K_s) \}_{s=1}^S
\end{equation}
and construct the PCHIP interpolant $\mathcal{F}_{\mathrm{PCHIP}}^{-1}(\cdot\ ; \mathcal{T})$. During online estimation, the normalized peak statistic $P_{\mathrm{obs}}^{(i)}$ of the $i$-th diffusion envelope is computed using the same normalization adopted when generating $\mathcal{T}$. The curvature estimate is then obtained by
\begin{equation}\label{eq:pchip_curvature_estimate}
\hat{K}_i = \mathcal{F}_{\mathrm{PCHIP}}^{-1}\left( P_{\mathrm{obs}}^{(i)}; \mathcal{T} \right).
\end{equation}
The corresponding distance estimate follows from the curvature definition as
\begin{equation}
\label{eq:distance_recovery}
\hat r_i
=
\frac{
D_{\mathrm a}^2
\cos^2\hat\theta_i
}{
2\lambda\hat K_i
}, 
\end{equation}
where \(\hat\theta_i\) and \(\hat K_i\) are obtained from
\eqref{eq:sine_domain_symmetry} and
\eqref{eq:pchip_curvature_estimate}, respectively. 

\begin{figure}[t]
    \centering
\includegraphics[width=1\linewidth]{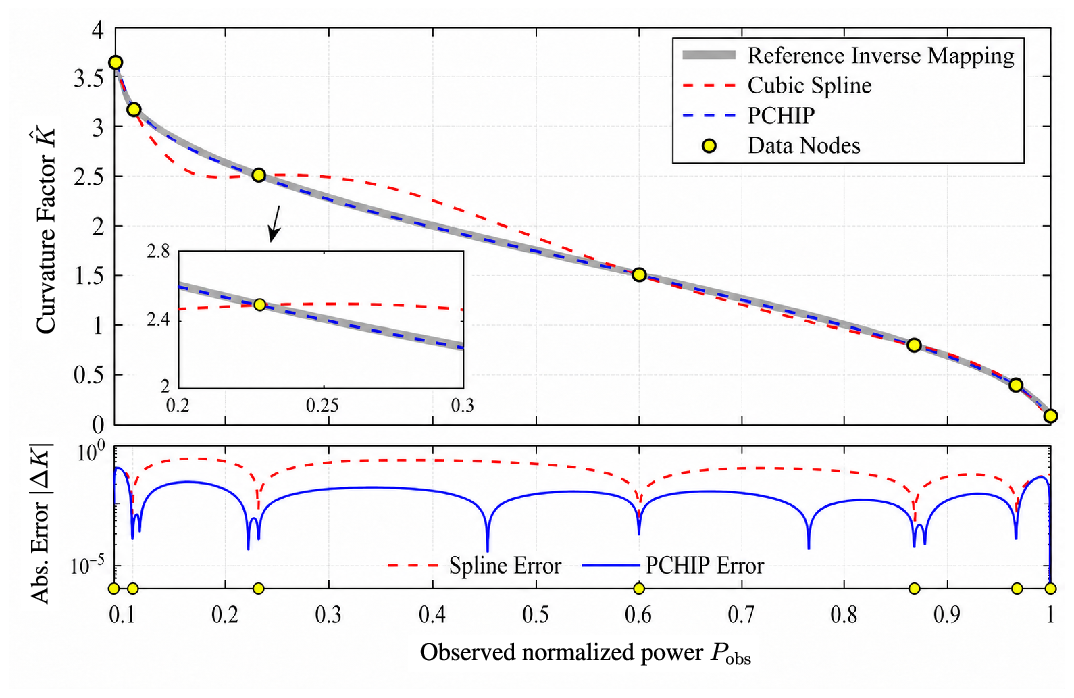}
\caption{Inverse power--curvature mapping and interpolation
error using cubic spline and PCHIP. Here, \(\hat K\) denotes
the interpolated curvature, and
\(\Delta K=\hat K-\mathcal F^{-1}(P_{\mathrm{obs}})\)
is the interpolation error.}
\label{fig:pchip_vs_spline}
\end{figure}

\textbf{Remark 2 (PCHIP-Based Distance Inversion):}
The inverse samples in $\mathcal{T}$  are generated over the monotone interval, $\mathcal D_{\mathrm{inv}}$, where $\mathcal F(K)$ is bijective by Lemma~2.  As shown in Fig.~\ref{fig:pchip_vs_spline}, PCHIP preserves the monotonicity of these samples and avoids the local overshoot that occurs with a conventional cubic spline \cite{e10}. Within the valid inversion interval, the mapping is generally better conditioned in its shorter range.  As $r_i$ increases, $K_i \to 0$,  and the power–curvature
mapping becomes progressively flatter.  Consequently, PCHIP can accurately approximate the inverse mapping, while the physical range resolution still decreases for sufficiently distant sources.

\subsection{Critical Distance Bound for Monotone Inversion}

Since the distance inversion in Lemma~2 is valid only over the monotone interval of $\mathcal F(K)$, we characterize the curvature boundary and translate it into a physical distance.

\textbf{Proposition 2 (Critical Distance for Monotone Inversion):}
For the power--curvature mapping $\mathcal F(K)$ in \eqref{eq:peak_power_response}, the first stationary point occurs at
$K_{\mathrm{cutoff}}\approx 3.65$. Hence, the bijective inversion is valid for an intra-site interference source satisfying
\begin{equation}
\label{eq:safe_distance}
r_i
\ge
R_{\mathrm C}(\theta_i)
=
\frac{
D_{\mathrm a}^2\cos^2\theta_i
}{
7.3\lambda
}.
\end{equation}

\begin{IEEEproof}
The boundary $K_{\mathrm{cutoff}}$ is determined by the first positive root of
$d\mathcal F(K)/dK=0$. Applying the quotient rule to
\eqref{eq:peak_power_response}, this condition is equivalent to
\begin{equation}
\label{eq:energy_balance}
K
\frac{d}{dK}
\left[
\mathcal C^2(\sqrt K)
+
\mathcal S^2(\sqrt K)
\right]
=
\mathcal C^2(\sqrt K)
+
\mathcal S^2(\sqrt K).
\end{equation}
Using the large-argument expansions of the Fresnel integrals,
\begin{equation}
\label{eq:asymptotic_fresnel}
\begin{cases}
\mathcal C(\sqrt K)
\approx
\dfrac{1}{2}
+
\dfrac{1}{\pi\sqrt K}
\sin\!\left(\dfrac{\pi}{2}K\right)
-
\dfrac{1}{\pi^2K^{3/2}}
\cos\!\left(\dfrac{\pi}{2}K\right), \\[6pt]
\mathcal S(\sqrt K)
\approx
\dfrac{1}{2}
-
\dfrac{1}{\pi\sqrt K}
\cos\!\left(\dfrac{\pi}{2}K\right)
-
\dfrac{1}{\pi^2K^{3/2}}
\sin\!\left(\dfrac{\pi}{2}K\right),
\end{cases}
\end{equation}
the right-hand side of \eqref{eq:energy_balance} can be approximated as
\begin{equation}
\label{eq:rhs_approx}
\mathcal C^2(\sqrt K)
+
\mathcal S^2(\sqrt K)
\approx
\frac{1}{2}
+
\frac{\sqrt 2}{\pi\sqrt K}
\sin\!\left(
\frac{\pi}{2}K-\frac{\pi}{4}
\right).
\end{equation}
Differentiating \eqref{eq:rhs_approx} and retaining the dominant phase term gives
\begin{equation}
\label{eq:lhs_approx}
\frac{d}{dK}
\left[
\mathcal C^2(\sqrt K)
+
\mathcal S^2(\sqrt K)
\right]
\approx
\frac{1}{\sqrt{2K}}
\cos\!\left(
\frac{\pi}{2}K-\frac{\pi}{4}
\right).
\end{equation}
Substituting \eqref{eq:rhs_approx} and \eqref{eq:lhs_approx} into
\eqref{eq:energy_balance} yields
\begin{equation}
\label{eq:transcendental_eq}
f(K)
=
\cos\!\left(
\frac{\pi}{2}K-\frac{\pi}{4}
\right)
-
\left[
\frac{1}{\sqrt{2K}}
+
\frac{2}{\pi K}
\sin\!\left(
\frac{\pi}{2}K-\frac{\pi}{4}
\right)
\right]
=
0 .
\end{equation}
The dominant cosine term gives the zeroth-order estimate
\begin{equation}
\label{eq:zeroth_order}
\frac{\pi}{2}K^{(0)}
-
\frac{\pi}{4}
=
\frac{3\pi}{2},
\quad
K^{(0)}=3.5 .
\end{equation}
Applying a first-order correction to \eqref{eq:transcendental_eq} gives
\begin{equation}
\label{eq:perturbation_correction}
\Delta K
\approx
-\frac{f(K^{(0)})}{f'(K^{(0)})}
=
\frac{2}{\pi}
\left[
\frac{1}{\sqrt{2K^{(0)}}}
+
\frac{2}{\pi K^{(0)}}
\sin\!\left(
\frac{3\pi}{2}
\right)
\right].
\end{equation}
With $K^{(0)}=3.5$, this gives $\Delta K\approx 0.125$, which provides an initial estimate near the first stationary point. Refining this estimate by solving
$d\mathcal F(K)/dK=0$ gives
$K_{\mathrm{cutoff}}\approx 3.65$.

From the definition
$K=D_{\mathrm a}^2\cos^2\theta_i/(2\lambda r_i)$, the condition
$K\le K_{\mathrm{cutoff}}$ is equivalent to
\eqref{eq:safe_distance}.
\end{IEEEproof}

\begin{figure}[t]
\centering
\includegraphics[width=1\linewidth]{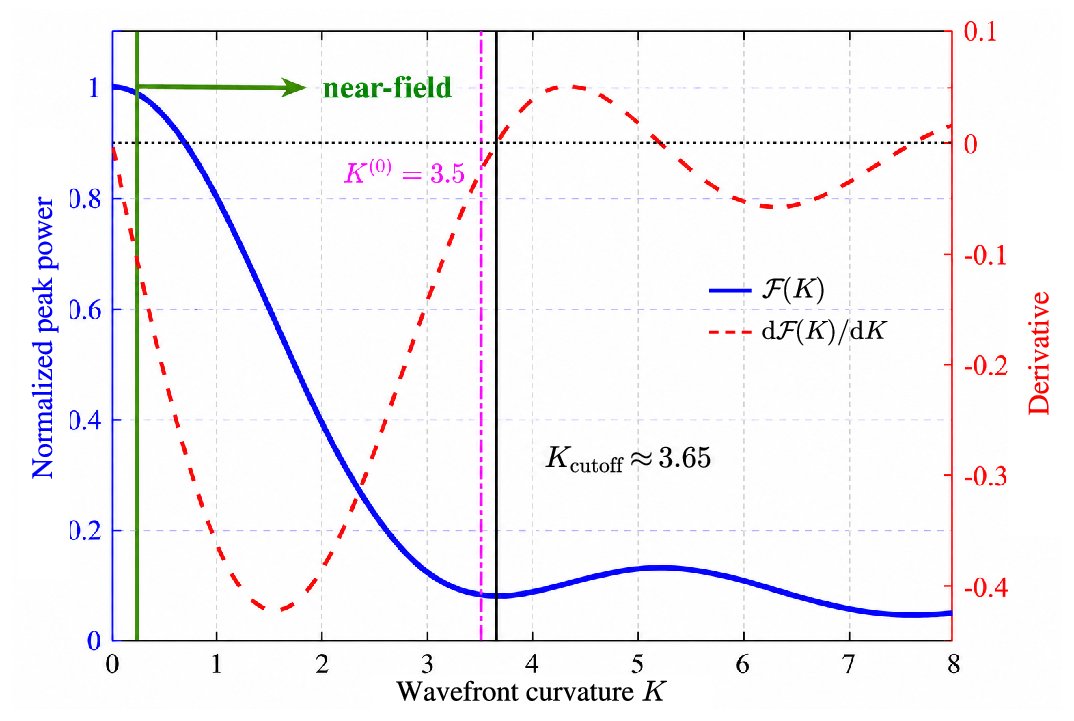}
\caption{Validation of the monotone inversion range and the corresponding critical distance bound.}
\label{fig:mse_breakdown}
\end{figure}

\textbf{Remark 3 (Physical Implication of the Critical Distance):}
Proposition~2 translates the monotone inversion range of $\mathcal F(K)$ into a physical distance bound. As shown in Fig.~\ref{fig:mse_breakdown}, $d\mathcal F(K)/dK$ remains negative from the Rayleigh reference point $K_{\mathrm{Ray}}=0.25$ to the first stationary point $K_{\mathrm{cutoff}}\approx 3.65$, which confirms the validity of the inversion over this curvature interval. Using the conventional Rayleigh distance
$R_{\mathrm{Ray}}=2D_{\mathrm a}^2/\lambda$ \cite{e7}, the critical distance in \eqref{eq:safe_distance} can be written as
\begin{equation}
\label{eq:critical_rayleigh_relation}
R_{\mathrm C}(\theta_i)
=
\frac{
R_{\mathrm{Ray}}\cos^2\theta_i
}{
4K_{\mathrm{cutoff}}
}
\approx
\frac{
R_{\mathrm{Ray}}\cos^2\theta_i
}{
14.6
}.
\end{equation}
Thus, the proposed inversion remains valid well inside the conventional Rayleigh distance. The factor $\cos^2\theta_i$ reflects the aperture projection effect. Specifically, off-broadside interference sources induce a smaller effective curvature, which enlarges the valid inversion range in terms of distance.

\section{Proposed PIN‑OMP Algorithm and Performance Analysis}\label{sec5}
This section presents the complete PIN-OMP algorithm for uplink channel estimation under hybrid-field
interference. PIN-OMP follows the dictionary-based sparse reconstruction framework in Section III, but replaces dense near-field angle--distance matching with the low-complexity parameter estimation developed in Section IV. The near-field interference basis reconstructed from the estimated parameters is first nulled. Then, residual far-field inter-site interference reconstruction and desired uplink support recovery are performed over projected far-field dictionaries.

\subsection{Proposed Algorithm}

PIN-OMP implements the framework in Section III with one modification: the dense angle--distance matching within the hybrid-field dictionary is replaced by the low-complexity interference-parameter estimation developed in Section IV. \(\widehat N_{\mathrm s}\), \(\widehat L_{\mathrm{inter}}\), and
\(\widehat L\) are determined online by APS-envelope detection and noise-adaptive
OMP stopping, with
\(\widehat K_{\mathrm{int}}
=\widehat N_{\mathrm s}+\widehat L_{\mathrm{inter}}\),
rather than using the generated path counts.

From the APS \(\mathcal P(\theta)\), the \(i\)-th dominant intra-site interference component is represented by the estimated parameters \((\hat\theta_i,\hat r_i)\). The near-field intra-site interference basis is constructed as

\begin{equation}
\label{eq:apn_intra_basis}
\hat{\mathbf A}_{\mathrm{intra}}
=
[
\mathbf b_{N_{\mathrm r}}(\hat\theta_1,\hat r_1),
\ldots,
\mathbf b_{N_{\mathrm r}}
(\hat\theta_{\widehat N_{\mathrm s}},
 \hat r_{\widehat N_{\mathrm s}})
].
\end{equation}

\begin{algorithm}[t]
\caption{PIN-OMP for Uplink Channel Estimation under Hybrid-Field Interference}
\label{alg:PIN_OMP}
\begin{algorithmic}[1]
 \STATE \textbf{Input:}
\(\mathbf Y\), \(\mathbf F_{\mathrm A}\), desired pilot vector
\(\mathbf x\), and the inverse map
\(\mathcal F_{\mathrm{PCHIP}}^{-1}(\cdot;\mathcal T)\).
\STATE \textbf{Initialization:} \(\Omega_{\rm inter}^{(0)}=\emptyset\), \(\Omega_{\rm UE}^{(0)}=\emptyset\),
 and \(\hat{\mathbf A}_{\rm intra}=[~]\).
\STATE Compute \(\hat{\mathbf h}_{\rm LS}\) using \eqref{eq:ls_estimation}, and compute \(\mathcal
 P(\theta)\) from \(\mathbf Y\).
 \STATE Detect the retained near-field diffusion envelopes
\(\{\mathcal E_i\}_{i=1}^{\widehat N_{\mathrm s}}\)
from \(\mathcal P(\theta)\).
\FOR{\(i=1,\ldots,\widehat N_{\mathrm s}\)}
\STATE Estimate \(\hat\theta_i\) using \eqref{eq:sine_domain_symmetry}.
\STATE Compute \(P_{\rm obs}^{(i)}\), obtain \(\hat K_i=
\mathcal F_{\mathrm{PCHIP}}^{-1}
\!\left(P_{\mathrm{obs}}^{(i)};\mathcal T\right)
\), and estimate \(\hat r_i\) using
 \eqref{eq:distance_recovery}.
\STATE Update
\(\hat{\mathbf A}_{\rm intra}
\leftarrow
[\hat{\mathbf A}_{\rm intra},
\mathbf b_{N_{\mathrm r}}(\hat\theta_i,\hat r_i)]\).
\ENDFOR
\STATE Compute \(\mathbf P_{\rm intra}^{\perp}\) using \eqref{eq:apn_intra_projection}.
\STATE Obtain
$\tilde{\mathbf Y}
=
\mathbf P_{\rm intra}^{\perp}\mathbf Y$ ,
 and 
$ \boldsymbol{\Phi}
=
\mathbf P_{\rm intra}^{\perp}\mathbf F_{\rm A}$. 
\STATE Set \(\mathbf Z^{(0)}=\tilde{\mathbf Y}\).
\STATE Run \(\widehat L_{\mathrm{inter}}\)-step OMP over \(\boldsymbol{\Phi}\) using the selection and residual update
rules in \eqref{eq:mf_omp_selection}--\eqref{eq:int_residual_update}, with \((\mathbf D,\mathbf
Y,\mathbf E)\) replaced by \((\boldsymbol{\Phi},\tilde{\mathbf Y},\mathbf Z)\). Obtain \(\Omega_{\rm
inter}\).
\STATE Form \(\hat{\mathbf A}_{\rm inter}=\mathbf F_{\rm A}(:,\Omega_{\rm inter})\) and \(\hat{\mathbf
 A}_{\rm int}\) using \eqref{eq:apn_composite_interference}.
\STATE Compute \(\mathbf P_{\rm int}^{\perp}\) using \eqref{eq:apn_total_projection}. 
\STATE  Obtain $\mathbf y_{\rm p}
=
\mathbf P_{\rm int}^{\perp}\hat{\mathbf h}_{\rm LS}$,
 and 
$\mathbf U_{\rm p}
=
\mathbf P_{\rm int}^{\perp}\mathbf F_{\rm A}$.
\STATE Run \(\widehat L\)-step OMP over \(\mathbf U_{\rm p}\) using
 \eqref{eq:ue_omp_selection}--\eqref{eq:ue_residual_update}. Obtain \(\hat{\Omega}_{\rm UE}\).
\STATE Form \(\hat{\mathbf A}_{\rm UE}\) using \eqref{eq:apn_ue_basis}.
\STATE Reconstruct \(\hat{\mathbf h}_{\rm UE}\) using \eqref{eq:apn_ue_reconstruction}.
\STATE \textbf{Output:} estimated desired uplink channel \(\hat{\mathbf h}_{\rm UE}\).
\end{algorithmic}
\end{algorithm}

The corresponding near-field nulling projection is
\begin{equation}
\label{eq:apn_intra_projection}
\mathbf P_{\rm intra}^{\perp}
=
\mathbf I_{N_{\mathrm r}}
-
\hat{\mathbf A}_{\rm intra}
\hat{\mathbf A}_{\rm intra}^{\dagger}.
\end{equation}

After near-field nulling, let $\boldsymbol{\Phi} = \mathbf{P}_{\mathrm{intra}}^{\perp}\mathbf{F}_{\rm A}$ denote the projected dictionary, $\tilde{\mathbf{Y}} = \mathbf{P}_{\mathrm{intra}}^{\perp}\mathbf{Y}$ denote the projected observation.
The residual far-field inter-site interference is reconstructed by OMP over \(\boldsymbol{\Phi}\). If the selected inter-site support is denoted by \(\Omega_{\rm inter}\), the corresponding inter-site basis is
$
\hat{\mathbf A}_{\rm inter}
=
\mathbf F_{\rm A}(:,\Omega_{\rm inter}).
$
The composite hybrid-field interference basis is then
\begin{equation}
\label{eq:apn_composite_interference}
\hat{\mathbf A}_{\mathrm{int}}
=
[
\hat{\mathbf A}_{\mathrm{intra}},
\hat{\mathbf A}_{\mathrm{inter}}
]
\in
\mathbb C^{
N_{\mathrm r}\times\widehat K_{\mathrm{int}}
},
\end{equation}
where \(\widehat K_{\rm int}\) is the selected number of reconstructed interference components. The composite interference-nulling projection is
\begin{equation}
\label{eq:apn_total_projection}
\mathbf P_{\rm int}^{\perp}
=
\mathbf I_{N_{\mathrm r}}
-
\hat{\mathbf A}_{\rm int}
\hat{\mathbf A}_{\rm int}^{\dagger}.
\end{equation}
The desired uplink support is detected over
\(\mathbf U_{\rm p}=\mathbf P_{\rm int}^{\perp}\mathbf F_{\rm A}\)
using the projected LS estimate
\(\mathbf y_{\rm p}=\mathbf P_{\rm int}^{\perp}\hat{\mathbf h}_{\rm LS}\).
Let \(\hat\Omega_{\rm UE}\) denote the detected support. The corresponding physical far-field basis is collected from the unprojected dictionary as
\begin{equation}
\label{eq:apn_ue_basis}
\hat{\mathbf A}_{\rm UE}
=
\mathbf F_{\rm A}(:,\hat\Omega_{\rm UE}).
\end{equation}
The desired uplink channel is reconstructed as
\begin{equation}
\label{eq:apn_ue_reconstruction}
\hat{\mathbf g}_{\rm UE}
=
\hat{\mathbf A}_{\rm UE}^{\dagger}\mathbf y_{\rm p},
\qquad
\hat{\mathbf h}_{\rm UE}
=
\hat{\mathbf A}_{\rm UE}\hat{\mathbf g}_{\rm UE}.
\end{equation}
The complete procedure is summarized in Algorithm~\ref{alg:PIN_OMP}. 

\subsection{Performance Analysis}

\subsubsection{Asymptotic Error Bounds for Distance Inversion}

We first characterize the estimation error in the distance inversion stage. Under ideal angle estimation, the normalized peak response of the $i$-th intra-site interference component is modeled as
\begin{equation}
\label{eq:power_observation_model}
P_{\mathrm{obs}}^{(i)}
=
\mathcal F(K_i)
+
w_P,
\end{equation}
where $w_P\sim\mathcal N(0,\sigma_P^2)$ denotes the equivalent power-domain perturbation, and $\mathcal F(K_i)$ is the on-axis power mapping defined in \eqref{eq:peak_power_response}.
Given the curvature-distance relation in \eqref{eq:distance_recovery}, 
we have
\begin{equation}
\frac{\partial r_i}{\partial K_i}
=
-\frac{r_i}{K_i}
=
-
\frac{
2\lambda r_i^2
}{
D_{\mathrm a}^2\cos^2\theta_i
}.
\end{equation}
Since the Fisher information for estimating $K_i$ from \eqref{eq:power_observation_model} is
$[\mathcal F'(K_i)]^2/\sigma_P^2$, the conditional CRB for estimating $r_i$ given the true angle $\theta_i$ is obtained as
\begin{equation}
\label{eq:crlb_ideal}
\mathrm{CRB}_{0}(r_i)
=
\left(
\frac{r_i}{K_i}
\right)^2
\frac{
\sigma_P^2
}{
[\mathcal F'(K_i)]^2
}
=
\frac{
4\lambda^2 r_i^4
}{
D_{\mathrm a}^4\cos^4\theta_i
}
\frac{
\sigma_P^2
}{
[\mathcal F'(K_i)]^2
}.
\end{equation}
The bound \eqref{eq:crlb_ideal} shows that distance inversion becomes more sensitive as the interference source moves farther from the array.
\begin{figure}[t]
    \centering
    \includegraphics[width=1\linewidth]{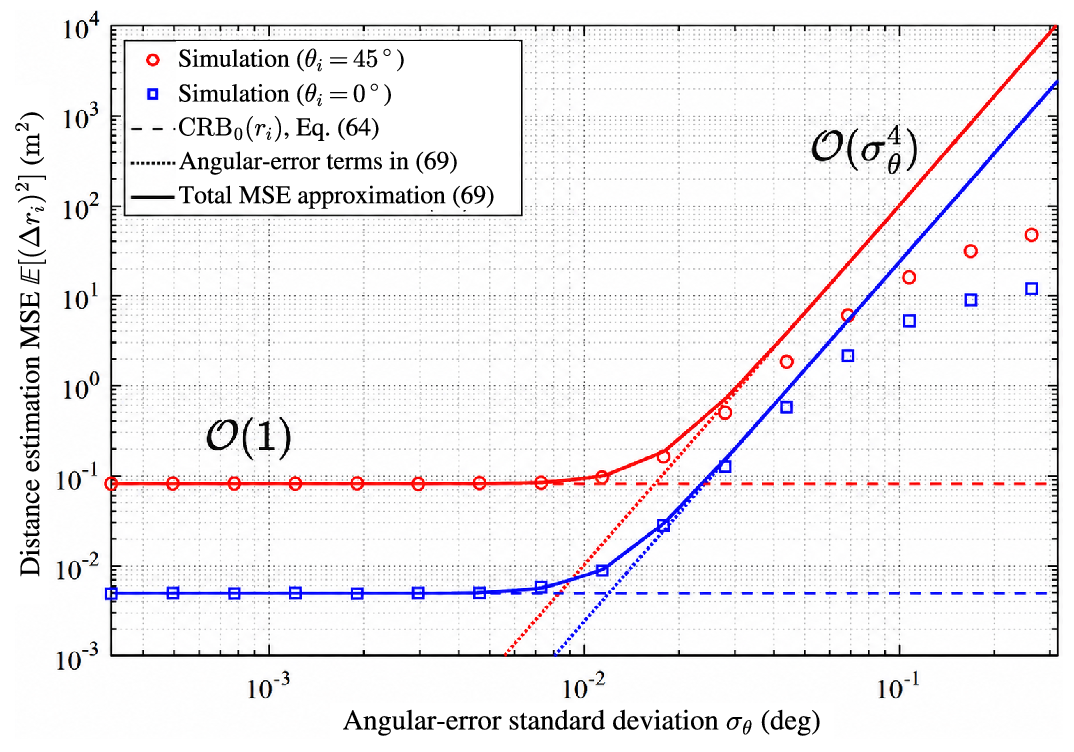}
  \caption{Error components of near-field interference distance
inversion. The annotations \(\mathcal O(1)\) and
\(\mathcal O(\sigma_\theta^4)\) indicate constant and quartic
contributions in \(\sigma_\theta\), respectively.}
    \label{fig:mse_breakdown_full}
\end{figure}
In practice, the angle estimation error is modeled as
\(\Delta\theta_i=\hat\theta_i-\theta_i
\sim\mathcal N(0,\sigma_\theta^2)\), where
\(\sigma_\theta\) is its standard deviation.
The distance error \(\Delta r_i=\hat r_i-r_i\)
can be locally approximated as

\begin{equation}
\label{eq:distance_error_linearization}
\Delta r_i
\approx
\frac{\partial r_i}{\partial K_i}\Delta K_i
+
\frac{\partial r_i}{\partial \theta_i}\Delta\theta_i,
\end{equation}
where
$\partial r_i/\partial\theta_i\!=\!-2r_i\tan\theta_i$.
Let $\bar{\mathcal P}(\theta,K)$ denote the normalized angular response, with
$\bar{\mathcal P}(\theta_i,K_i)=\mathcal F(K_i)$, and define
\begin{equation}
\label{eq:second_derivative_power}
\bar{\mathcal P}_{\theta\theta}(K_i)
\triangleq
\left.
\frac{
\partial^2\bar{\mathcal P}(\theta,K)
}{
\partial\theta^2
}
\right|_{\theta=\theta_i,K=K_i}.
\end{equation}
Since $\theta_i$ is the symmetry center of the diffusion envelope, the first-order angular derivative vanishes. A second-order expansion gives
\begin{equation}
\label{eq:taylor_expansion_bound}
P_{\mathrm{obs}}^{(i)}
\approx
\mathcal F(K_i)
+
\mathcal F'(K_i)\Delta K_i
+
\frac{1}{2}
\bar{\mathcal P}_{\theta\theta}(K_i)
(\Delta\theta_i)^2 .
\end{equation}
Combining \eqref{eq:taylor_expansion_bound} with \eqref{eq:power_observation_model} yields
\begin{equation}
\label{eq:delta_K}
\Delta K_i
=
\frac{
w_P
-
\frac{1}{2}
\bar{\mathcal P}_{\theta\theta}(K_i)
(\Delta\theta_i)^2
}{
\mathcal F'(K_i)
}.
\end{equation}

Assuming that $w_P$ and $\Delta\theta_i$ are independent, and using
$\mathbb E[(\Delta\theta_i)^4]=3\sigma_\theta^4$, the distance MSE is obtained as
\begin{equation}
\label{eq:generalized_mse_final}
\begin{aligned}
\mathbb E[(\Delta r_i)^2]
\approx&
\left(
\frac{r_i}{K_i}
\right)^2
\frac{
\sigma_P^2
}{
[\mathcal F'(K_i)]^2
}
+
4r_i^2\tan^2\theta_i\,\sigma_\theta^2
\\
&+
\frac{
3r_i^2
}{
4K_i^2
}
\left(
\frac{
\bar{\mathcal P}_{\theta\theta}(K_i)
}{
\mathcal F'(K_i)
}
\right)^2
\sigma_\theta^4 .
\end{aligned}
\end{equation}
\textbf{Remark 4 (Error Decoupling in Distance Inversion):}
The first term in \eqref{eq:generalized_mse_final} corresponds to the angle-error-free CRB in \eqref{eq:crlb_ideal}. The second term is the direct geometric propagation of the angular error through the factor $\cos^2\theta_i$ in the curvature--distance relation. The third term is caused by evaluating the angular response at $\hat\theta_i$ instead of the true angle $\theta_i$, and therefore depends on the fourth moment of the angular error. As shown in Fig.~\ref{fig:mse_breakdown_full}, the power-domain perturbation dominates when the angular estimate is accurate, whereas the angular-error terms become dominant as the angular mismatch increases.

\subsubsection{Conditional HCRB for Channel Reconstruction}

We derive a conditional HCRB \cite{e9} for desired uplink channel reconstruction, conditioned on the reconstructed interference basis $\hat{\mathbf A}_{\mathrm{int}}$ and a given UE support $\Omega_{\mathrm{UE}}$. Let $\mathbf A_{\mathrm{UE}}=\mathbf F_{\rm A}(:,\Omega_{\mathrm{UE}})$ denote the corresponding far-field UE basis. The LS estimate can be written as
\begin{equation}
\label{eq:hcrb_obs_model}
\hat{\mathbf h}_{\mathrm{LS}}
=
\hat{\mathbf A}_{\mathrm{int}}\boldsymbol{\eta}
+
\mathbf A_{\mathrm{UE}}\boldsymbol{\alpha}
+
\tilde{\mathbf n},
\end{equation}
where $\boldsymbol{\eta}$ contains deterministic interference coefficients, $\boldsymbol{\alpha}$ contains random UE path gains, and
$\tilde{\mathbf n}\sim
\mathcal{CN}(\mathbf 0,\sigma_{\tilde n}^2\mathbf I_{N_{\mathrm r}})$.
For notational simplicity, the normalization factor in (2) is absorbed into the effective gain vector \(\boldsymbol\alpha\). We model
\(\boldsymbol\alpha\sim
\mathcal{CN}(\mathbf 0,\sigma_\alpha^2\mathbf I_L)\),
where \(\sigma_\alpha^2\) is chosen consistently with the adopted channel-power normalization. Define
$\gamma_{\mathrm{reg}}\triangleq\sigma_{\tilde n}^2/\sigma_{\alpha}^2$.
Let $\mathbf{A}_{\mathrm{joint}} \triangleq \bigl[\hat{\mathbf{A}}_{\mathrm{int}}, \mathbf{A}_{\mathrm{UE}}\bigr]$ and $\mathbf{c}_{\mathrm{joint}} \triangleq \bigl[\boldsymbol{\eta}^T, \boldsymbol{\alpha}^T\bigr]^T$. The corresponding hybrid information matrix (HIM) is
\begin{equation}
\label{eq:him_block}
\mathbf{J}_{\mathrm{H}}
=
\frac{1}{\sigma_{\tilde{n}}^2}
\mathbf{A}_{\mathrm{joint}}^H
\mathbf{A}_{\mathrm{joint}}
+
\begin{bmatrix}
\mathbf{0} & \mathbf{0} \\
\mathbf{0} & \sigma_{\alpha}^{-2} \mathbf{I}_{L}
\end{bmatrix}
=
\begin{bmatrix}
\mathbf{J}_{11} & \mathbf{J}_{12} \\
\mathbf{J}_{21} & \mathbf{J}_{22}
\end{bmatrix},
\end{equation}
where the sub-blocks are given by $\mathbf{J}_{11} = \sigma_{\tilde{n}}^{-2} \hat{\mathbf{A}}_{\mathrm{int}}^H \hat{\mathbf{A}}_{\mathrm{int}}$, $\mathbf{J}_{12} = \mathbf{J}_{21}^H = \sigma_{\tilde{n}}^{-2} \hat{\mathbf{A}}_{\mathrm{int}}^H \mathbf{A}_{\mathrm{UE}}$, and $\mathbf{J}_{22} = \sigma_{\tilde{n}}^{-2} \mathbf{A}_{\mathrm{UE}}^H \mathbf{A}_{\mathrm{UE}} + \sigma_{\alpha}^{-2} \mathbf{I}_{L}$.
Assuming that $\hat{\mathbf A}_{\mathrm{int}}$ has full column rank, the deterministic nuisance parameter $\boldsymbol{\eta}$ is eliminated through the Schur complement. The HCRB for $\boldsymbol{\alpha}$ is then given by
\begin{equation}
\begin{aligned}
\label{eq:hcrb_alpha}
\mathbf C_{\mathrm{HCRB}}(\boldsymbol{\alpha})
&= \left(\mathbf{J}_{22} - \mathbf{J}_{21} \mathbf{J}_{11}^{-1} \mathbf{J}_{12}\right)^{-1}\\
&= \sigma_{\tilde n}^{2}
\left(
\mathbf A_{\mathrm{UE}}^H
\mathbf P_{\mathrm{int}}^{\perp}
\mathbf A_{\mathrm{UE}}
+
\gamma_{\mathrm{reg}}
\mathbf I_{L}
\right)^{-1},
\end{aligned}
\end{equation}
where $\mathbf P_{\mathrm{int}}^{\perp}$ is defined in \eqref{eq:apn_total_projection}.
Since
$\mathbf h_{\mathrm{UE}}=\mathbf A_{\mathrm{UE}}\boldsymbol{\alpha}$,
the corresponding conditional HCRB is given by
\begin{equation}
\label{eq:mse_far_field_hcrb}
\begin{aligned}
\mathbb{E}
&\left[
\left\|
\hat{\mathbf{h}}_{\mathrm{UE}} - \mathbf{h}_{\mathrm{UE}}
\right\|_2^2
\right]
\ge
\operatorname{tr}
\left(
\mathbf{A}_{\mathrm{UE}}
\mathbf{C}_{\mathrm{HCRB}}(\boldsymbol{\alpha})
\mathbf{A}_{\mathrm{UE}}^H
\right)
\\
&=
{\sigma_{\tilde{n}}^2}
\operatorname{tr}
\left(
\mathbf{A}_{\mathrm{UE}}
\left(
\mathbf{A}_{\mathrm{UE}}^H
\mathbf{P}_{\mathrm{int}}^{\perp}
\mathbf{A}_{\mathrm{UE}}
+
\gamma_{\mathrm{reg}}
\mathbf{I}_{L}
\right)^{-1}
\mathbf{A}_{\mathrm{UE}}^H
\right).
\end{aligned}
\end{equation}
For the special case where
$\mathbf A_{\mathrm{UE}}^H\mathbf A_{\mathrm{UE}}
=
\mathbf I_{L}$,
let
$\mathbf G_{\Omega}
\triangleq
\mathbf A_{\mathrm{UE}}^H
\mathbf P_{\mathrm{int}}^{\perp}
\mathbf A_{\mathrm{UE}}$,
and let $\lambda_k$ denote its $k$-th eigenvalue. Then \eqref{eq:mse_far_field_hcrb} reduces to
\begin{equation}
\label{eq:mse_far_field_hcrb_eig}
\mathbb E
\left[
\left\|
\hat{\mathbf h}_{\mathrm{UE}}
-
\mathbf h_{\mathrm{UE}}
\right\|_2^2
\right]
\ge
\sigma_{\tilde n}^{2}
\sum_{k=1}^{L}
\frac{1}{\lambda_k+\gamma_{\mathrm{reg}}}.
\end{equation}
This bound shows that a small $\lambda_k$, corresponding to a UE-channel direction that is strongly attenuated by the interference-nulling projection, increases the reconstruction lower bound. 
Hence, the eigenvalues of \(\mathbf G_{\Omega}\) characterize the overlap between the desired UE subspace and the reconstructed interference subspace. 

\subsection{Computational Complexity}

Table~\ref{tab:complexity_comparison} summarizes the dominant computational complexity of PIN-OMP and the benchmark estimators. Let $G_{\mathrm u}$ denote the 1D angular scanning grid size for APS construction, $G_{\mathrm f}$ denote the far-field dictionary size, and $G_{\rm n}=G_\theta G_r$ denote the polar-domain dictionary size. Only leading-order terms are retained; scalar operations such as PCHIP evaluation are omitted.

\begin{table}[t]
\centering
\caption{Dominant Computational Complexity Comparison}
\label{tab:complexity_comparison}
\renewcommand{\arraystretch}{1.25}
\begin{tabular}{l c}
\toprule
\textbf{Algorithm} & \textbf{Dominant Computational Complexity} \\
\midrule
Standard LS
& $\mathcal{O}(N_{\mathrm r}\tau)$ \\
Blind LMMSE
& $\mathcal{O}(N_{\mathrm r}^3+N_{\mathrm r}^2\tau)$ \\
1D FF-OMP
& $\mathcal{O}(K_{\mathrm{int}}N_{\mathrm r}G_{\rm f}\tau+LN_{\mathrm r}G_{\rm f})$ \\
Hybrid-Field OMP
& $\mathcal{O}\!\left(K_{\mathrm{int}}N_{\mathrm r}\tau(G_{\rm n}+G_{\rm f})+LN_{\mathrm r}G_{\rm f}\right)$ \\
\textbf{Proposed PIN-OMP}
& $\mathcal{O}\!\left(N_{\mathrm r}[\tau(G_{\rm u} + L_{\mathrm{inter}}G_{\rm f}) + LG_{\rm f}]\right)$ \\
\bottomrule
\end{tabular}
\end{table}

For PIN-OMP, the dominant cost comes from APS construction, residual inter-site interference extraction, and UE support recovery. Under direct scanning, APS construction requires $\mathcal{O}(G_{\rm u}N_{\mathrm r}\tau)$, while an FFT implementation reduces this term to $\mathcal{O}(\tau G_{\rm u}\log G_{\rm u})$. The residual far-field inter-site interference reconstruction over the projected far-field dictionary requires $\mathcal{O}(L_{\mathrm{inter}}N_{\mathrm r}G_{\rm f}\tau)$, and the null-space OMP for UE support recovery requires $\mathcal{O}(LN_{\mathrm r}G_{\rm f})$. The low-rank projection and pseudo-inverse operations introduce an additional overhead of
$\mathcal{O}(N_{\mathrm r}K_{\mathrm{int}}G_{\rm f}+N_{\mathrm r}K_{\mathrm{int}}^2)$,
which is omitted from Table~\ref{tab:complexity_comparison} since $K_{\mathrm{int}}\ll G_{\rm f}$ in sparse interference settings.
Compared with Hybrid-Field OMP, PIN-OMP replaces the exhaustive 2D near-field dictionary search over $G_{\rm n}$ atoms with 1D APS construction and low-complexity parameter inversion. Hence, the complexity reduction mainly comes from eliminating the near-field angle--distance grid search while retaining 1D sparse recovery for the inter-site components and the desired uplink channel.

\section{Simulation Results}\label{sec6}

In this section, numerical examples are presented to evaluate the proposed PIN-OMP algorithm for desired uplink channel estimation under hybrid-field interference. Unless otherwise specified, the XL-array BS is equipped with a ULA of $N_{\mathrm r}=128$ receive antennas and operates at $f_{\mathrm c}=30$ GHz, corresponding to $\lambda=1$ cm. The antenna spacing is set to $d=\lambda/2$. The pilot length is $\tau=40$, and all pilot sequences are generated from normalized 16-QAM symbols. The results are averaged over 2000 independent Monte Carlo trials.
The desired uplink channel is modeled as a far-field multipath channel with $L=3$ paths. The UE angular center is uniformly generated over $[-60^\circ,60^\circ]$. The intra-site interference consists of $N_{\mathrm s}=3$ near-field components located at
$\theta_{\mathrm{intra}}\in\{-30^\circ,0^\circ,30^\circ\}$ and
$r_{\mathrm{intra}}\in\{12,10,9\}$ m. Their INR values are set to $\{18,20,22\}$ dB, respectively. The inter-site interference contains $L_{\mathrm{inter}}=2$ far-field components with angles $\{-45^\circ,45^\circ\}$ and INR $10$ dB. The desired signal-to-noise ratio (SNR) is $-15$ dB.
The following benchmark schemes are considered.

\begin{itemize}
    \item \textbf{Standard LS:} The pilot-correlated LS estimator without interference suppression.

    \item \textbf{Blind LMMSE \cite{e_mmse}:} An LMMSE estimator using a statistically averaged UE spatial covariance matrix as a blind prior.

    \item \textbf{1D FF-OMP \cite{lee2016omp}:} A far-field OMP estimator that extracts interference and UE supports over the 1D angular dictionary.

    \item \textbf{Hybrid-Field OMP \cite{e_hf}:}
    A sparse-recovery baseline that performs OMP over a hybrid-field dictionary, consisting of a 2D near-field angle--distance dictionary and a 1D far-field angular dictionary.

    \item \textbf{Conditional HCRB reference \eqref{eq:mse_far_field_hcrb}:} The conditional bound derived in Section~V, evaluated with the true interference basis and the given UE support.
\end{itemize}

\begin{figure}[t]
    \centering
    \includegraphics[width=0.95\linewidth]{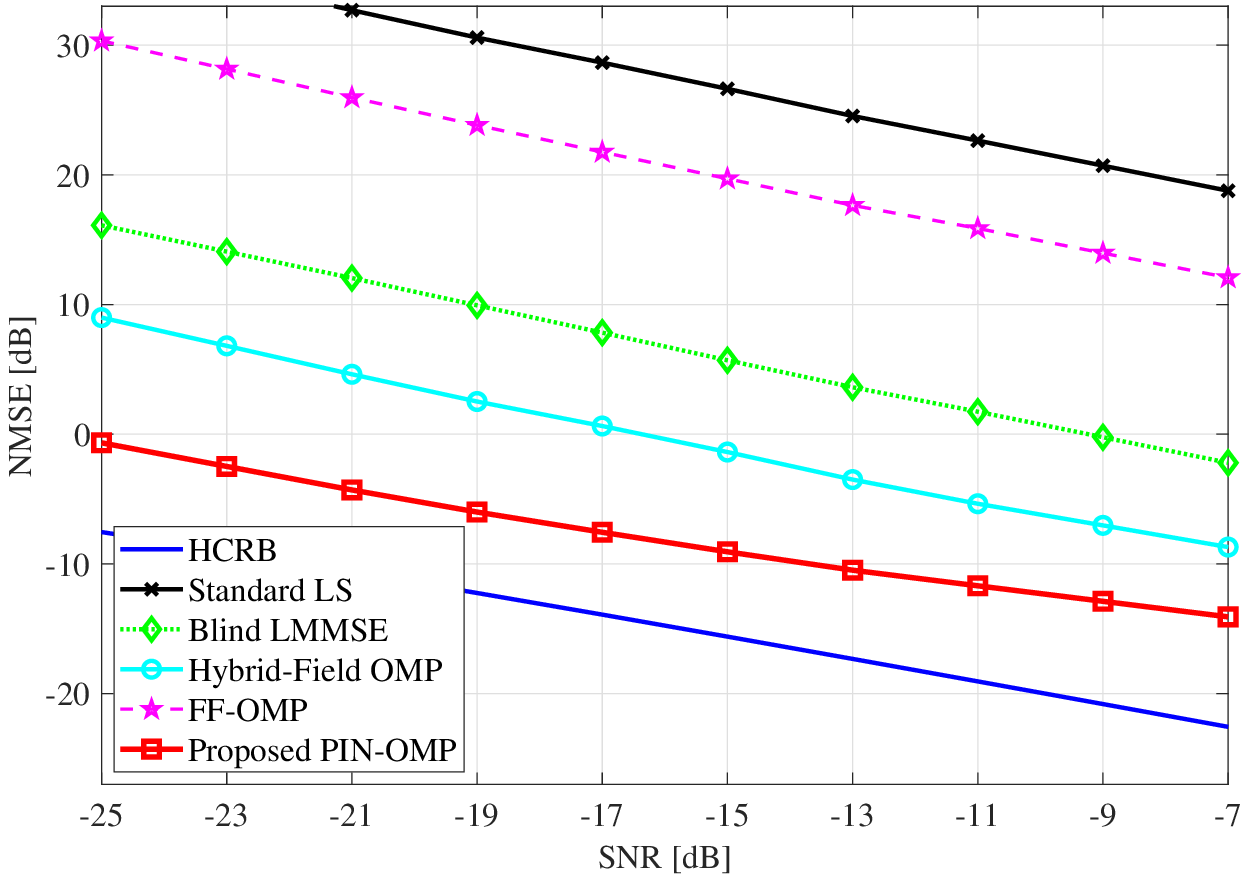}
\caption{NMSE of desired uplink channel estimation versus SNR under hybrid-field interference, with $N_{\mathrm r}=128$ and $\tau=40$.}
\label{fig:nmse_snr_tau40}
\end{figure}

The normalized mean square error (NMSE) is adopted as the performance metric and is defined as
\begin{equation}
\label{eq:simulation_nmse}
\mathrm{NMSE}
=
\mathbb E
\left[
\frac{
\|\hat{\mathbf h}_{\mathrm{UE}}-\mathbf h_{\mathrm{UE}}\|_2^2
}{
\|\mathbf h_{\mathrm{UE}}\|_2^2
}
\right],
\end{equation}
which is the expectation of the squared relative estimate error.

\begin{figure}[t]
    \centering
    \includegraphics[width=0.95\linewidth]{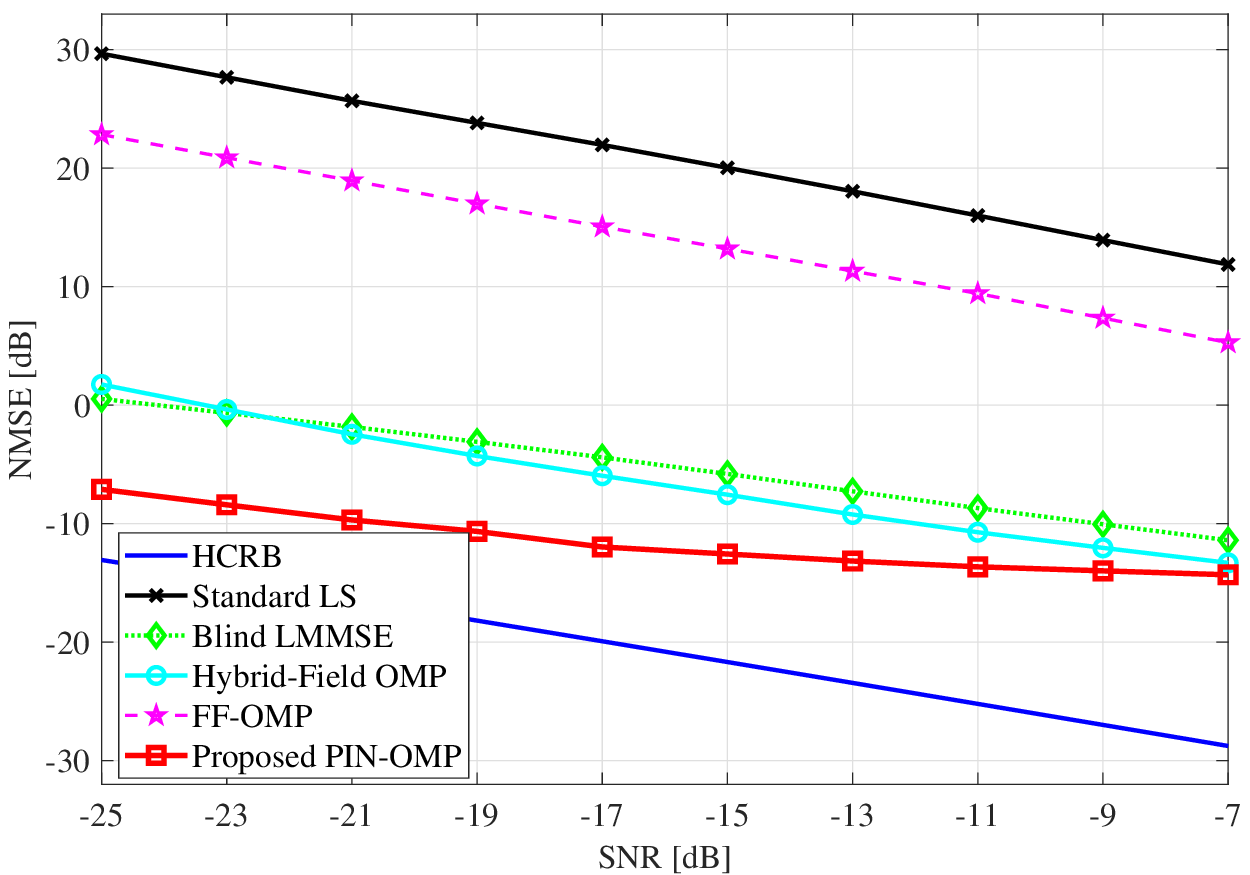}
\caption{NMSE of desired uplink channel estimation versus SNR under hybrid-field interference, with $N_{\mathrm r}=128$ and $\tau=200$.}
\label{fig:nmse_snr_tau200}
\end{figure}

\subsection{Runtime Comparison}

Table~\ref{tab:runtime_comparison} compares the online runtime
of the considered schemes. Replacing the exhaustive near-field
dictionary search with APS-based parameter estimation reduces the
average runtime from \(107.509\,\mathrm{ms}\) for Hybrid-Field OMP
to \(11.287\,\mathrm{ms}\) for PIN-OMP, corresponding to an
approximately one-order-of-magnitude (\(9.53\times\)) reduction.
Blind LMMSE and 1D FF-OMP are included as computational references.

\begin{table}[t]
\caption{Runtime comparison among different schemes}
\label{tab:runtime_comparison}
\centering
\footnotesize
\renewcommand{\arraystretch}{1.08}
\begin{tabular*}{\columnwidth}{
@{\extracolsep{\fill}}lcc}
\toprule
Scheme
& Mean \(\pm\) Std. (ms)
& Median (ms) \\
\midrule
Blind LMMSE
& \(0.807 \pm 0.130\)
& \(0.763\) \\
1D FF-OMP
& \(12.379 \pm 1.616\)
& \(11.937\) \\
Hybrid-Field OMP
& \(107.509 \pm 8.410\)
& \(104.182\) \\
Proposed PIN-OMP
& \(11.287 \pm 1.734\)
& \(10.903\) \\
\bottomrule
\end{tabular*}
\end{table}

\subsection{Channel Estimation Performance Versus SNR}

Figs.~\ref{fig:nmse_snr_tau40} and \ref{fig:nmse_snr_tau200} compare the SNR-domain NMSE performance of the desired uplink channel estimate under two representative pilot lengths, $\tau=40$ and $\tau=200$, respectively. In both cases, standard LS suffers from severe hybrid-field interference because it does not suppress either near-field intra-site interference or far-field inter-site interference. FF-OMP improves over LS by exploiting the far-field angular sparsity, but its 1D far-field dictionary cannot accurately represent the intra-site interference components. Blind LMMSE suppresses part of the interference through statistical filtering, while its performance depends on the accuracy of the residual covariance estimate. Hybrid-Field OMP further reduces the NMSE by using a hybrid-field dictionary, but remains limited by finite grid resolution and dictionary coherence. In contrast, PIN-OMP consistently achieves the lowest NMSE among all practical estimators over the simulated SNR range. This gain is obtained by reconstructing the dominant intra-site interference, extracting the residual inter-site interference after near-field nulling, and estimating the UE support over the projected far-field dictionary. PIN-OMP also remains closest to the conditional HCRB reference, and the remaining gap is mainly caused by interference-parameter estimation errors and finite dictionary resolution in the complete algorithm.

\begin{figure}[t]
    \centering
    \includegraphics[width=0.95\linewidth]{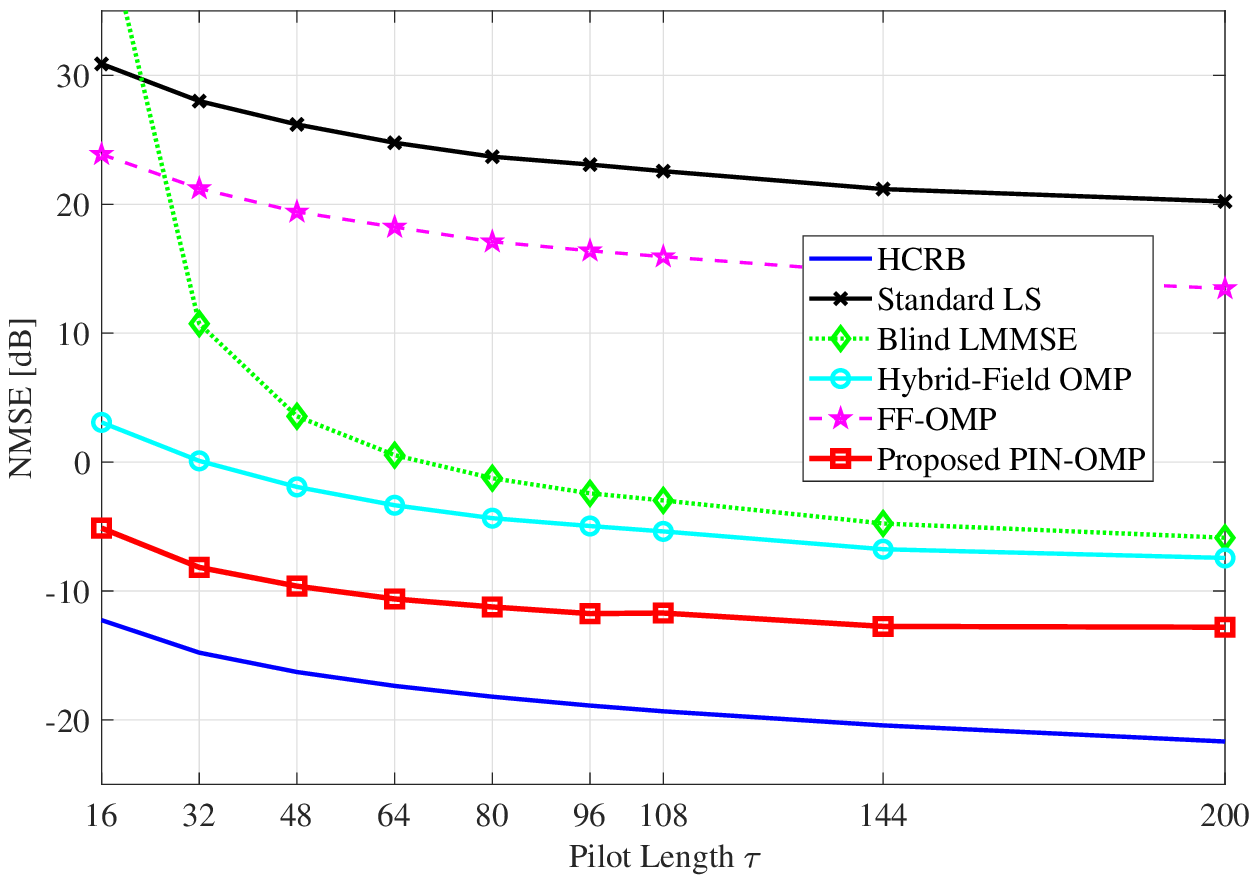}
    \caption{NMSE of desired uplink channel estimation versus pilot length $\tau$ under hybrid-field interference, with $N_{\mathrm r}=128$ and SNR $=-15$ dB.}
    \label{fig:nmse_pilot}
\end{figure}
\subsection{Effect of Pilot Length}

Fig.~\ref{fig:nmse_pilot} evaluates the NMSE of the desired uplink channel estimate versus the pilot length $\tau$ at a fixed SNR of $-15$ dB. Increasing $\tau$ improves all schemes because pilot correlation reduces the effective noise. Standard LS remains interference-limited since it does not suppress hybrid-field interference, while FF-OMP is constrained by its far-field representation of near-field intra-site interference. Blind LMMSE benefits from longer pilots but remains sensitive to the accuracy of the residual covariance estimate. Hybrid-Field OMP improves over FF-OMP by using a hybrid-field dictionary, but its performance is still affected by finite grid resolution and dictionary coherence. PIN-OMP achieves the lowest NMSE among all practical estimators over the tested pilot lengths. The gain is more evident in the short-pilot regime, where the reconstruction of the dominant intra-site interference and subsequent residual inter-site extraction provide a more stable interference-nulling subspace. As $\tau$ increases, PIN-OMP follows the same decreasing trend as the conditional HCRB.

\begin{figure}[!t]
    \centering
    \includegraphics[width=0.95\linewidth]{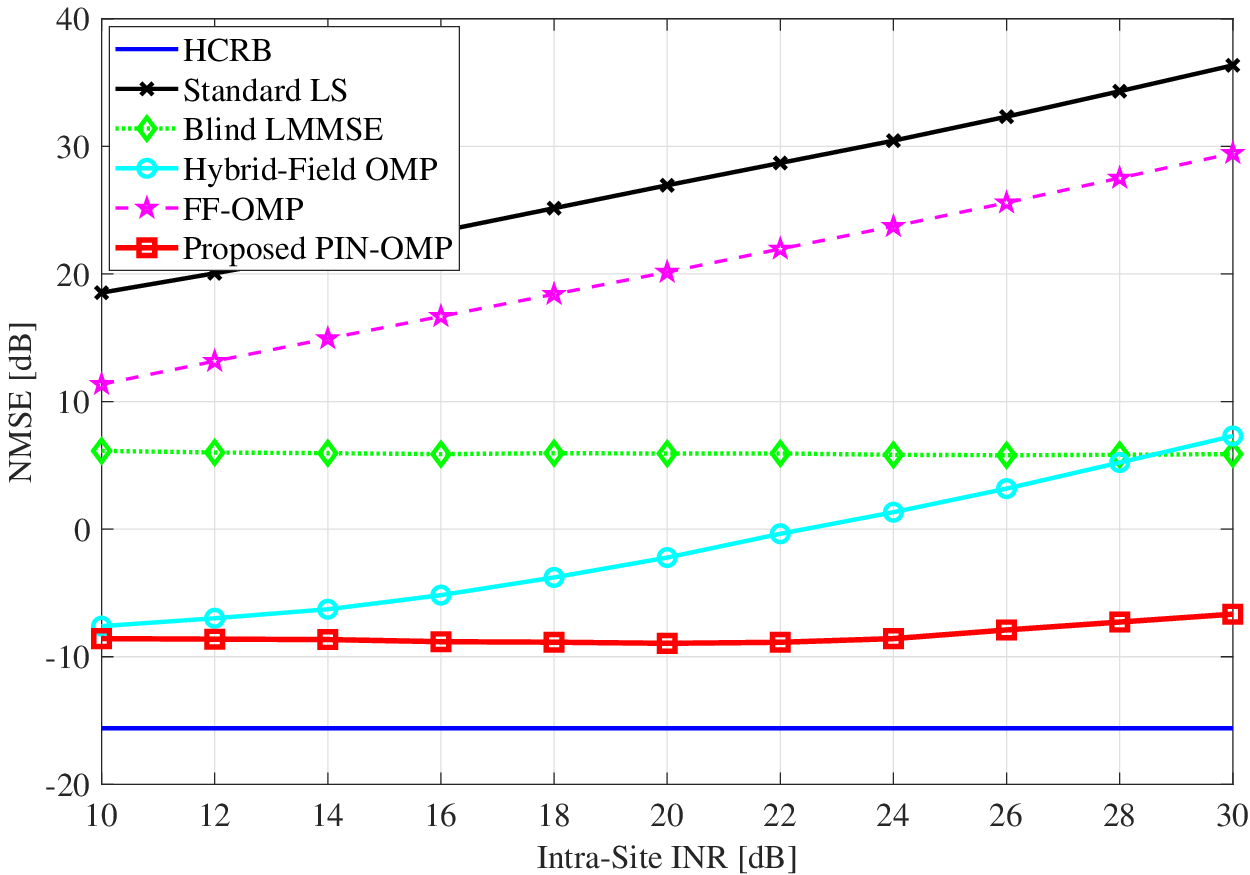}
    \caption{Desired uplink channel NMSE versus intra-site INR under hybrid-field interference, with $N_{\mathrm r}=128$, $\tau=40$, and SNR $=-15$ dB.}
    \label{fig:nmse_inr}
\end{figure}

\subsection{Effect of Intra-Site INR}

Fig.~\ref{fig:nmse_inr} evaluates the NMSE of the desired uplink channel estimate versus the intra-site INR at a fixed SNR of $-15$ dB and pilot length $\tau=40$. As the intra-site power increases, standard LS degrades rapidly because the pilot-correlated LS estimate is increasingly dominated by the interference components. FF-OMP also deteriorates with INR, since the 1D far-field dictionary cannot represent the near-field interference subspace. Blind LMMSE remains less sensitive to the INR variation, but provides limited suppression because it relies on a statistical UE prior rather than explicit hybrid-field interference reconstruction. Hybrid-Field OMP improves over FF-OMP by using a hybrid-field dictionary, but its performance becomes less stable in the high-INR regime due to finite grid resolution and dictionary coherence. In contrast, PIN-OMP maintains the lowest NMSE among all practical estimators over most of the considered INR range. This gain comes from the reconstruction of the dominant near-field intra-site interference and the subsequent extraction of residual far-field inter-site interference before UE support recovery. At high INR, the gap between PIN-OMP and the conditional HCRB increases, since residual errors in the reconstructed intra-site interference subspace are amplified by stronger interference.
\begin{figure}[!t]
    \centering
    \includegraphics[width=0.95\linewidth]{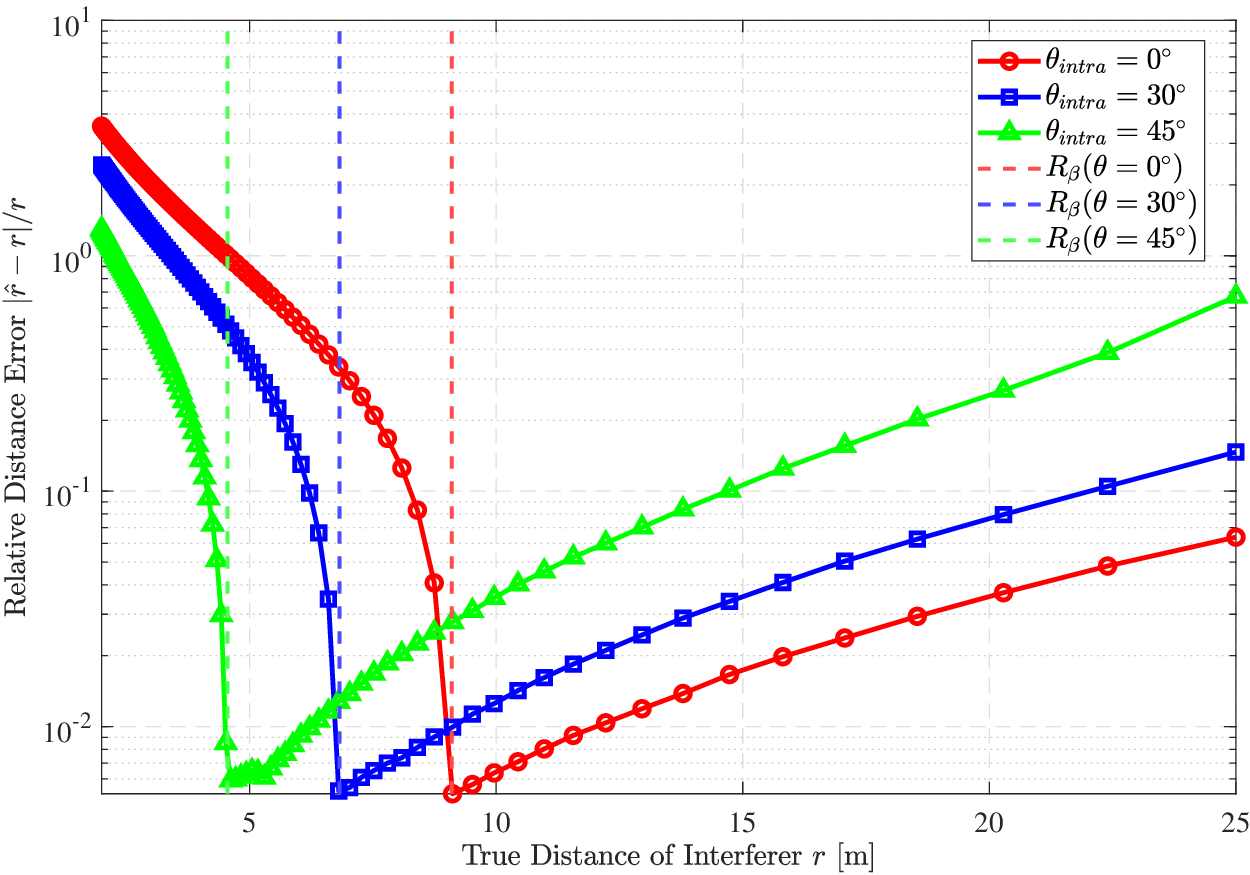}
    \caption{Relative distance estimation error of intra-site interference versus true distance for different interference angles. The dashed lines denote the implemented PCHIP inversion boundaries.}
    \label{fig:distance_error_curve}
\end{figure}
\begin{figure}[!t]
    \centering
    \includegraphics[width=0.95\linewidth]{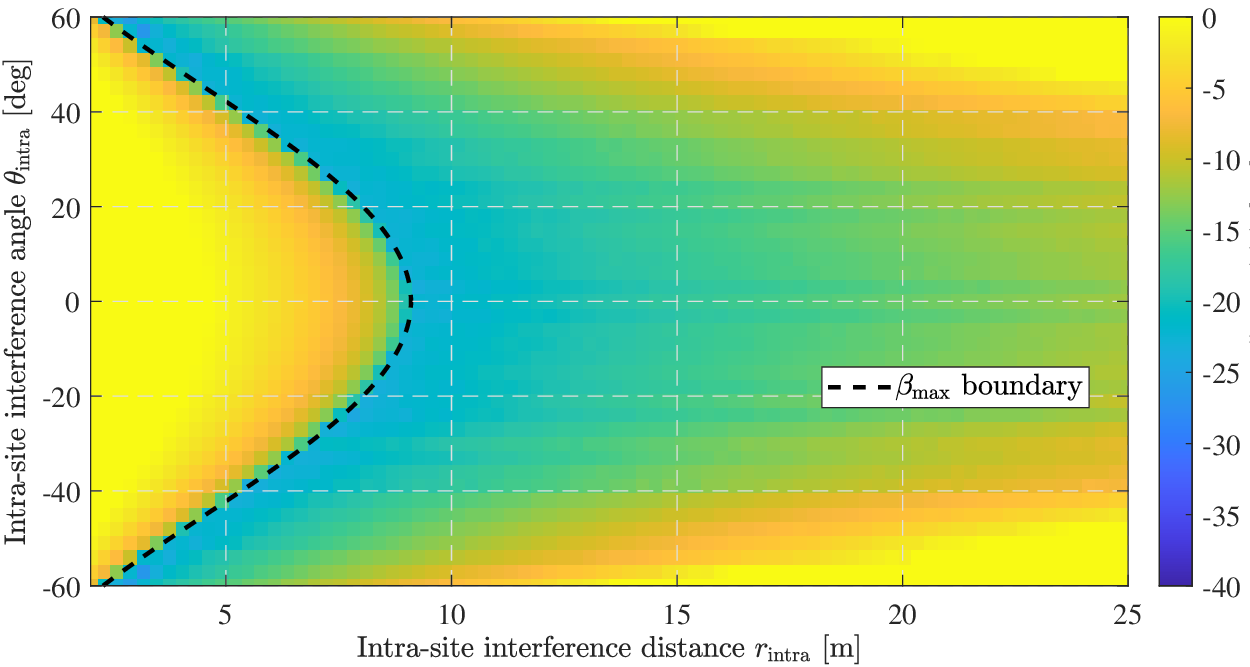}
\caption{Relative distance estimation error map over intra-site interference distance and angle, with $N_{\mathrm r}=128$, $\tau=40$, SNR $=-10$ dB, and INR $=20$ dB. The dashed curve denotes the implemented PCHIP inversion boundary.}
    \label{fig:distance_error_map}
\end{figure}

\subsection{Feasibility Region of Analytical Distance Inversion}

For numerical implementation, the curvature factor is parameterized as
\begin{equation}
\label{eq:beta_definition_sim}
\beta
\triangleq
\sqrt K
=
\frac{
D_{\mathrm a}|\cos\theta_{\mathrm{intra}}|
}{
\sqrt{2\lambda r_{\mathrm{intra}}}
}.
\end{equation}
The PCHIP lookup table used in the simulations is constructed over
$\beta\in[10^{-2},\beta_{\max}]$, where $\beta_{\max}=1.50$. Hence, the tabulated inversion covers curvature values up to
$K_{\mathrm{tab}}=\beta_{\max}^2$. The corresponding implemented inversion boundary is
\begin{equation}
\label{eq:implemented_beta_boundary}
R_{\beta}(\theta_{\mathrm{intra}})
=
\frac{
D_{\mathrm a}^2\cos^2\theta_{\mathrm{intra}}
}{
2\lambda\beta_{\max}^2
}.
\end{equation}
For $r_{\mathrm{intra}}<R_{\beta}(\theta_{\mathrm{intra}})$, the corresponding curvature exceeds the tabulated range, and the distance inversion becomes sensitive to observation perturbations. Therefore, $R_{\beta}(\theta_{\mathrm{intra}})$ represents the practical operating boundary of the PCHIP inversion.

Fig.~\ref{fig:distance_error_curve} evaluates the relative distance estimation error
$|\hat r_{\mathrm{intra}}-r_{\mathrm{intra}}|/r_{\mathrm{intra}}$
for representative intra-site interference angles. For each angle, the error is large when the source lies before the corresponding implemented boundary, and decreases rapidly once $r_{\mathrm{intra}}$ enters the tabulated inversion region. The boundary moves toward shorter distances as $|\theta_{\mathrm{intra}}|$ increases, which follows from the projected aperture factor $\cos^2\theta_{\mathrm{intra}}$ in \eqref{eq:implemented_beta_boundary}. This behavior is consistent with the curvature-dependent distance inversion in \eqref{eq:distance_recovery}.

Fig.~\ref{fig:distance_error_map} further shows the relative distance error over the two-dimensional $(r_{\mathrm{intra}},\theta_{\mathrm{intra}})$ plane. The low-error region is primarily on the valid side of the implemented PCHIP boundary, whereas larger errors occur at shorter distances and near broadside, where the effective curvature is stronger. These results characterize the practical feasibility region of the analytical distance estimation used for near-field intra-site interference reconstruction in PIN-OMP.
\vspace{-0.2cm}

\section{Conclusion}\label{sec7}

In this paper, we have investigated uplink channel estimation under hybrid-field interference, which consists of near-field intra-site interference and far-field inter-site interference. To suppress them, we have developed a low-complexity interference reconstruction framework. Specifically, we have first estimated interference parameters based on the angular diffusion of the dominant near-field interference in the far-field angular spectrum.
Based on the estimated parameters, we have developed the PIN-OMP algorithm to null the reconstructed near-field interference basis, recover residual far-field inter-site interference, and reconstruct the desired uplink channel from the interference-suppressed observation. We have also derived a physical critical-distance bound for monotone distance inversion and a conditional HCRB for the desired channel reconstruction limit, together with a dominant complexity analysis. Numerical results have verified that PIN-OMP achieves better NMSE performance than representative baseline schemes under different SNRs, pilot lengths, and intra-site INR values.
\vspace{-0.2cm}

\bibliographystyle{IEEEtran}
\bibliography{main.bib}

\end{document}